\documentclass[useAMS,usenatbib,usedcolumn]{mn2e}
\input psfig.sty
\usepackage{epsfig}
\usepackage{amsmath,amssymb}
\usepackage{psfig}
\usepackage{float}
\usepackage{graphicx}
\usepackage{epstopdf}
\usepackage{url}
\usepackage{natbib}
\newcommand{\Mpc} {{\,\rm Mpc}}
\newcommand{\kpc} {{\,\rm kpc}}

\newcommand{\kms}{{\,\rm {km\,s^{-1}} }}

\renewcommand{\vec}[1]{{\mathbf #1}}
\newcommand{\mpch}{\>h^{-1}{\rm {Mpc}}}   

\def \apj  {ApJ}
\def \apjs  {ApJS}
\def \apjl  {ApJL}
\def \prd {Phy.Rev.D}
\def\pr{Phys.Rev}
\def \mnras {MNRAS}
\def \sci {Science}
\def \aap {A\&A}
\def \jcap {JCAP}
\def \etal {et~al.~}\def \chisq  {\ifmmode  \chi^2   \else  $\chi^2$  \fi}  
\def \spose#1{\hbox  to 0pt{#1\hss}}  
\def \lta{\mathrel{\spose{\lower 3pt\hbox{$\sim$}}\raise  2.0pt\hbox{$<$}}}
\def \gta{\mathrel{\spose{\lower  3pt\hbox{$\sim$}}\raise 2.0pt\hbox{$>$}}}

\def \kms {\ifmmode  \,\rm km\,s^{-1} \else $\,\rm km\,s^{-1}  $ \fi }
\def \kpc {\ifmmode  {\rm kpc}  \else ${\rm  kpc}$ \fi  }  
\def \Msun {\ifmmode M_{\odot} \else $M_{\odot}$ \fi} 
\def \hMsun {\ifmmode h^{-1}\,\rm M_{\odot} \else $h^{-1}\,\rm M_{\odot}$ \fi}

\def \LCDM {\ifmmode \Lambda{\rm CDM} \else $\Lambda{\rm CDM}$ \fi}
\def \sig8 {\ifmmode \sigma_8 \else $\sigma_8$ \fi} 
\def \OmegaM {\ifmmode \Omega_{\rm M} \else $\Omega_{\rm M}$ \fi} 
\def \OmegaL {\ifmmode \Omega_{\rm \Lambda} \else $\Omega_{\rm \Lambda}$\fi} 
\def \Deltavir {\ifmmode \Delta_{\rm vir} \else $\Delta_{\rm vir}$ \fi}
\def \rhocrit {\ifmmode \rho_{\rm crit} \else $\rho_{\rm crit}$ \fi}
\def \rhou {\ifmmode \rho_{\rm u} \else $\rho_{\rm u}$ \fi}
\def \zc {\ifmmode z_{\rm c} \else $z_{\rm c}$ \fi}

\def \rhos {\ifmmode \rho_{\rm s} \else $\rho_{\rm s}$ \fi} 
\def \rs {\ifmmode r_{\rm s} \else $r_{\rm s}$ \fi} 
\def \cvir {\ifmmode c_{\rm vir} \else $c_{\rm vir}$ \fi} 
\def \Rvir {\ifmmode r_{\rm vir} \else $R_{\rm vir}$ \fi}
\def \Vvir {\ifmmode V_{\rm  vir} \else  $V_{\rm vir}$  \fi} 
\def \Mvir {\ifmmode M_{\rm  vir} \else $M_{\rm  vir}$ \fi}  
\def \Nvir {\ifmmode N_{\rm  vir} \else $N_{\rm  vir}$ \fi}  
\def \Jvir {\ifmmode J_{\rm vir} \else $J_{\rm vir}$ \fi} 
\def \Evir {\ifmmode E_{\rm vir} \else $E_{\rm vir}$ \fi} 
\def \lam {\ifmmode \lambda  \else $\lambda$ \fi} 
\def \lamp {\ifmmode \lambda^{\prime} \else $\lambda^{\prime}$  \fi} 
\def \Vmax {\ifmmode V_{\rm  max} \else  $V_{\rm max}$  \fi}

\title[Structure formation in WDM]{Structure formation in warm dark matter cosmologies Top-Bottom Upside-Down}
\author[S. Paduroiu \etal]
{Sinziana Paduroiu$^{1}$\thanks{E-mail: sinziana.paduroiu@unige.ch}, Yves Revaz$^{2}$, Daniel Pfenniger$^1$\\
$^1$Geneva Observatory, University of Geneva, CH-1290 Sauverny, Switzerland\\
$^2$Laboratoire d'Astrophysique, \'Ecole Polytechnique F\'ed\'erale de Lausanne (EPFL), 1290 Sauverny, Switzerland
}
\date{\today}
\begin{document}
\maketitle

\begin{abstract}
  The damping on the fluctuation spectrum and the presence of thermal velocities as properties of warm dark matter particles like
  sterile neutrinos imprint a distinct signature found from the structure formation mechanisms to the internal structures of halos.
  Using warm dark matter simulations we explore these effects on the structure formation for different particle energies and we find
  that the formation of structure is more complex than originally assumed, a combination of top-down collapse and hierarchical (bottom-up)
  clustering on multiple scales. The degree on which one scenario is more prominent with respect to the other depends
  globally on the energy of the particle and locally on the morphology and architecture of the analyzed region. The presence of shells and caustics in
  warm dark matter halos is another important effect seen in simulations. Furthermore we discuss the impact of
  thermal velocities on the structure formation from theoretical considerations as well as from the analysis of the simulations. We
  re-examine the assumptions considered when estimating the velocity dispersion for warm dark matter particles that have been
  adopted in previous works for more than a decade and we give an independent estimation for the velocities. We identify some
  inconsistencies in previous published results.  The relation between the warm dark matter particle mass and its corresponding
  velocity dispersion is strongly model dependent, hence the constraints on particle mass from simulation results are weak.  Finally,
  we review the technical difficulties that arise in warm dark matter simulations along with possible improvements of the methods.

\end{abstract}


\section{Introduction}

Independent studies and observations of both small and large scale structure are presently challenging the otherwise widely embraced
CDM model. The so-called missing satellites problem \citep[e.g.][]{klypin99, moore99}, where observations of galaxies do not map the
abundance of substructures that are produced in CDM cosmologies is a serious drawback of the model.  Furthermore, at smaller scales,
the density profiles of galaxies show large cores \citep[e.g.][]{deBlok01, salucci2012, deNaray} that have not been reproduced by
the simulations.  The failure to replicate in CDM simulations pure bulgeless galaxies which are observed in an important fraction
\citep{Kormendy10} adds to the problem.

While several recipes have been proposed in the attempt of ameliorating these issues \citep[e.g.][]{nfw96, martizzi, Mashchenko08,
  Pontzen12}, most of them introducing baryonic physics processes, current studies conclude that even including repeated baryonic
outflows, large cored galaxies are not found in the simulations \citep{Marinacci}, although this is still highly debated in the
literature.

The aforementioned situations, where the CDM model proves deficient in explaining the observations, are demanding further
investigation. The WDM models, with sterile neutrinos leading as most probable particle candidates have been well studied and
discussed in the literature in the past thirty years with an increased interest in the last few years (see the highlights of Daniel
Chalonge workshops and Colloquiums 2011-2013 for latest developments in the WDM field \citep[]{Ch11, Ch12}).

It has been shown recently \citep{Destri:2012yn} that modeling the quantum pressure of fermionic particles \citep{Weinberg,
  Muccione} on the other hand, one can reproduce the expected cores in dwarf galaxies, known to be dark matter dominated.

Moreover, the recent detection \citep{Esra, Boy14} of a 3.55\,keV unidentified emission line both on the data from XMM spectrum of
galaxy clusters and Chandra can be a hint of sterile neutrino decay.

Since particles in warm dark matter models have different intrinsic properties from the cold dark matter particle candidates, the
effect of these particle on structure formation and evolution is expected to be qualitatively different on both large and small
scales.

Notwithstanding the difficulties in modeling properly the neutrino particle, several attempts \citep[e.g.][]{Colombi, Bode01,
  Maccio2012, Kamada2013, PLM} have been conducted with a successful outcome in solving some of the cases where CDM comes to an
impasse.  While the methods of CDM simulations have been accurately improved over the last decade, the WDM simulations encounter a
number of difficulties in accurately describing the effect of such particles on both large and small scales. In addition to the
resolution limitations that are met in the CDM case as well, WDM particles like neutrinos, for example, have a phase space density
limit, a Fermi-Dirac distribution and a thermal velocity dispersion. Referring merely to sterile neutrinos, these particles decouple
whilst relativistic.

The effects of the initial velocities of the warm dark matter particle are expected to manifest themselves on small scale structure
formation. The free streaming exponentially dampens the power spectrum of density fluctuations such that very few structures are formed below the damping scale. Conservation of the fine grained phase space density is expected to set a maximum density
that cannot be exceeded during the formation of structures with collisionless particles. For a fermionic WDM particle, we can
crudely define the coarse-grained phase-space density $Q \equiv \rho/\sigma^{3}$, where $\rho$ is the density and $\sigma$ is the
velocity dispersion.  This definition is only a good approximation for locally isotropic velocities where the density of particles
is not strongly varying.

Different numerical approaches have been used to address the impact of warm dark matter particles thermal velocities.  Since the
numerical resolution is strongly limited with respect to the physics, one knows that the phase space distribution sampling is anyway
poor in space as well as in velocity space. The best compromise is to imprint the physical particle velocity to the simulation
particle, as common practice in galactic dynamics.  The particle limited sampling is not a sufficient reason to entirely drop the velocity sampling by neglecting the thermal velocities as done in some previous works, while keeping nonetheless the power spectrum cutoff implied by a non-zero thermal velocity \citep[e.g.][]{PLM,GOV}.  Nor is the fact that for some dark matter particles the thermal velocities are comparable or smaller than the bulk Zel'dovich velocities.  Even though it has been considered difficult to
prescribe accurately initial thermal velocities in dark matter simulations, the importance of using them has been emphasized in
previous studies like \cite{Colombi}, \cite{Bode01} and \cite{Melott}.

In the absence of a tested universal mechanism of production for the warm dark matter particle, the relation between the particle
mass and its corresponding thermal velocity is strongly dependent on the specific model adopted. The widely used formula for
generating velocities \citep{Bode01} which sets such a relation is based on an assumption that overestimates the number of
species at decoupling and in effect underestimates the value of thermal velocities. We will take the opportunity to discuss these
assumptions and we will also provide a method for estimating thermal velocities for fermionic, maxwellian and bosonic particles in
both relativistic and non-relativistic regimes based on a different set of premises that takes quantum physics into consideration.

Several analyses of warm dark matter simulations in the keV range conclude that the formation of structure is hierarchical, like in
cold dark matter simulations. Traditionally, top-down structure formation means forming chronologically the biggest structures first
and the smaller ones later, while bottom-up or hierarchical structure formation means the reverse scale order, making it difficult
to describe a scenario in which both coexist. In fact it is well known since, e.g., \cite{Lin} and \cite{Zel} that typical
structure formation proceeds in time first along pancakes, then filaments and then halos, mixing the large and small spatial scales
at all times with different proportions.  If we use this terminology in a broader sense, top-down describes the dominant long range
effects on structure formation: sheets collapsing into filaments, collapsing into halos.  Bottom-up hierarchical structure
formation, on the other hand, describes dominant short scale effects, mergers of both early formed and later formed halos.  We
will examine how both of these mechanisms of structure formation show up in the warm dark matter simulations presented here.

Additional constraints coming from peculiar features may be considered.  In cold dark matter models, during the hierarchical
evolution caustics are being wrapped inside earlier generations of the merging history, making them invisible in some cases even at
high resolutions. However, \cite{Cooper10} show using cold dark matter simulations that accretion mechanisms of stars
and dark matter clumps and disruption of the latter can produce concentering shells that resemble those observed in NGC 7600.  In
the warm dark matter simulations, as we will see, the shells and caustics are more visible, especially at high redshift, where the
top-down formation occurs.

Constraints on the mass of a warm dark matter particle from Lyman-$\alpha$ Forest, cosmic weak lensing, gamma-ray bursts,
etc. \citep[e.g.][]{Markovic, Souza} give a lower limit in the few keV range. To study the effects of warm dark matter on 
structure formation, we have, however, explored a larger mass interval, focusing on the region where these effects are more
prominent while fairly balanced by the resolution.

The paper is structured as follows: in Section 2 we explain the theoretical reasons for using the thermal component of velocities in
the simulations. Subsection 2.2.1 shows how the common used formula for generating velocities in warm dark matter simulations
\citep[]{Bode01} is conjectured from hypothetical assumptions. Section 2.2.2 presents a different approach in estimating thermal
velocities from the particle mass. In Section 2.3 we discuss some of the inconsistencies found in previous studies. Section 3
describes the parameters used in our simulations while Section 4 shows the results found from analyzing the simulations. At last, we
present our conclusions in Section 5.

\section{Initial conditions of cosmological simulations}
\subsection{Velocities in the initial conditions of cosmological simulations}

The initial conditions of most CDM and WDM cosmological simulations have often initial thermal velocity taken as
strictly zero, with the argument that at the finite initial redshift the thermal velocity of CDM particles is small in
regard of the bulk flow following from Zel'dovich's prescription.  We argue below that this practice is numerically inconsistent
with the actual problem of describing a collisionless fluid of finite phase space density $f$.  For structure formation,
the distribution of the dark matter particles in velocity space is most important, as stressed by \cite{Colombi}.
 
Indeed, integrating Newton's equation of motion for a set of particles in a force field $\vec g$ is equivalent to solving in a
Lagrangian way with discrete mass particles the collisionless Boltzmann equation with the characteristics method.  In conventional
notations the Eulerian description of the phase space volume conservation reads,
\begin{equation}
\label{eq:phase1}
{\frac{\partial{f }}{\partial {t}}} 
+ {\vec v} \cdot {\frac{\partial{f}}{\partial{\vec x}}}
+ {\vec g} \cdot {\frac{\partial{f}}{\partial{\vec v}}}= 0,
\end{equation}
where $ {\vec g}$ is the force field.  The mass density $\rho$ is the projection of the phase space density on velocity space:
\begin{equation}
\label{eq:phase2}
\rho(\vec x, t) = \int d^3v \, f(\vec x, \vec v, t).
\end{equation}

The mass density $\rho $ generates the force field $\vec g$ by Newton's gravity.  In a cosmological setup the mean density $\rho_0$
is subtracted,
\begin{eqnarray}
\label{eq:phase3}
\vec g &=& G \int d^3x'\, \left[\rho(\vec x',t)-{\bar\rho}_0(t)\right]
              { \vec x-\vec x' \over |\vec x-\vec x'|^3} \cr
      & =& 
           G \int d^3x'\,d^3v\, \left[f(\vec x',\vec v,t)-{\bar f}_0(\vec v,t)\right] 
             { \vec x-\vec x' \over |\vec x-\vec x'|^3} .
\end{eqnarray}
So in this context using vanishing small thermal velocity already poses a consistency problem since $f$ is of the form
$\delta(\vec v-\vec v_0(\vec x))\rho(\vec x)$.  This implies representing the system with a diverging $f$ in a vanishing fraction of
the phase space volume, in other words, mass belongs only to an infinitesimally thin 3D sheet in 6D phase space.  As $f$ is
conserved along a characteristics, it implies that this singularity must persist at all subsequent times.  Methods to conserve
arbitrarily high phase space density have been set up \citep{Abel, Hahn}, but this is not necessarily sufficient.  Ideally a physically
sound solution $f_0$ to this Boltzmann-Poisson system should not be numerically sensitive to the initial condition
discretization. In practice it is known that the gravitational $N$-body problem is exponentially sensitive to perturbations
\citep{Miller}, so the best that can be expected in such simulations is that over an ensemble of simulations with identical
statistical initial conditions, results follow a reproducible statistical distribution. Detailed evolution of particles is sensitive
to perturbations, but the average evolution of an ensemble of particles is predictable.

When $f_0$ is finite and differentiable everywhere, in other words when $f_0$ mathematically exists, the sound situation that should
be used, a slight variation of $f_0$, a fluctuation, will also follow the same set of equations, so, writing $f = f_0+f_1$, $\vec g
= \vec g_0 + \vec g_1$, where $f_1$ and $\vec g_1$ are the differences between the reference $f_0$ and the perturbed solution $f$,
and using the fact that $f_0$ is a solution of the system, we obtain the exact equations for the differences $f_1$, and $\vec
g_1$:
\begin{equation}
\label{eq:phase4}
\frac{\partial{f_1 }}{\partial{t}} 
+ {\vec v} \cdot {\frac{\partial{f_1}}{\partial{\vec x}}} + {\vec g_0} \cdot {\frac{\partial{f_1}}{\partial{\vec v}}} =- 
 {\vec g_1} \cdot {\frac{\partial{f_0}}{\partial{\vec v}}} - {\vec g_1} \cdot {\frac{\partial{f_1}}{\partial{\vec v}}}
\end{equation}
\begin{equation}
\label{eq:phase5}
\vec g_1 = G \int d^3x'\,d^3v\, \, f_1(\vec x',\vec v,t)  
{(\vec x-\vec x')\over |\vec x-\vec x'|^3} 
\end{equation}
So we see that $f_1$ follows the same left-hand side equation than $f_0$ in the unperturbed field $\vec g_0$, except that now the
right-hand side contains a source term whose first term ${\vec g_1} \cdot {\frac{\partial{f_0}}{\partial{\vec v}}}$ is the product
of the force fluctuations $\vec g_1$ times the gradient of the original $f_0$ in velocity space.  Thus a vanishing zero thermal
velocity for a set of particles supposed to represent a physical $f_0$ is not only suspicious since it corresponds to a delta
function in velocity space, but also because the gradient $ {\frac{\partial {f_0}}{\partial {\vec v}}}$ is at least as singular as
$f_0$.  In other words a zero initial thermal width is inconsistent with the initial assumptions, and susceptible to arbitrary
strong amplification of perturbations, since the variational equations contain a diverging source term to first order when the
initial thermal velocity is small.  The only possibility to cancel this diverging term is either to have vanishing force
fluctuations ${\vec g_1}$, which is exceptional when $f_1$ is non-zero, or that ${\vec g_1}$ is orthogonal to
$ {\frac{\partial {f_0}}{\partial {\vec v}}}$, which is also exceptional.  The second order term
${\vec g_1} \cdot {\frac{\partial{f_1}}{\partial{\vec v}}}$ on the right-hand side can cancel the first term only when $f_1$ is
proportional and opposite to $f_0$, which is also exceptional.  In summary dealing with diverging $f_0$ is inconsistent with the
implicit assumption of regularity of the mathematical problem.

It is instructive to compare how simulating collisionless fluids is differently approached in the fields of stellar and
galactic dynamics.  While in cosmology the physical collisionless fluids is assumed to consist of elementary particles,
in galactic dynamics the fluid is composed of stars.  In both cases the numerical simulation particles are order of
magnitudes more massive than the physical particles.  In the CDM context a simulation particle velocity is seen as
representing the bulk flow of a large ensemble of CDM particles, explaining why zero velocity dispersion has been
assumed. In galactic dynamics in contrast it is well known that doing so would immediately cause huge gravitational 
instabilities, and that the correct way to perform collisionless galaxy simulations is to ascribe to the simulation
particles the same velocity dispersion as the stars.  This is also required for respecting the virial theorem. While the
use of velocities has been shown \citep{Melott} to be of importance for the CDM simulations, in WDM it becomes even more
relevant \citep{Colombi} since the particles do have intrinsic non-zero velocity dispersion.  Collisionless fluids are
particle mass agnostic, when the particle mass starts to be important is when the 2-body relaxation time is shorter than
the system age. Recalling the Chandrasekhar 2-body relaxation time  in an arbitrarily large homogeneous medium 
\citep{Chandrasekhar, Henon}, 
\begin{equation}   
 \tau_\mathrm{rel}={v^3 \over 8\pi G^2 m \rho \ln N},
\label{eq:relax}
\end{equation}  
where $v$ is the velocity dispersion of $N$ particles of mass $m$ with average density $\rho$, it is obvious that this 
relaxation time is proportional to  $v^3$, $1/m$,  and $1/\ln N$, so is exactly zero when $v=0$.  Thus CDM simulations with 
initial zero $v$ are technically collisional until numerical noise heats the particles to larger $v$.

From a completely different side of physics, the maximal phase space density constraint set by Heisenberg's inequality,
\begin{equation}
\label{eq:heis}
\Delta{x}\Delta{p_x}\geq\frac{\hbar}{2}
\end{equation}
gives a minimum particle velocity dispersion, taking $\Delta{x}=n^{-1/3}$ and $m$ the particle mass,
\begin{equation}
\label{eq:vmin}
\Delta{v}\geq\frac{\hbar}{2}\frac{n^{1/3}}{m} \approx 0.003  \left[1\,\mathrm{keV}/c^2 \over m \right] 
      \left[n \over 1\, \mathrm{cm}^{-3}\right]^{1/3} \mathrm{km/s} \ .
\end{equation}
Taking a velocity dispersion lower than this value violates Heisenberg inequality, while taking it only slightly larger means that
the particles behaviour is governed by quantum physics, not classical mechanics as assumed in all cosmological $N$-body simulations.

In summary, adopting even a slight non-zero velocity dispersion in cosmological simulation is certainly a safer and more correct
physical assumption than taking strictly cold initial conditions associated with a mathematically singular and inconsistent state
leading to situations not under control on the numerical viewpoint due to singular phase space density.

\subsection{Thermal velocities of warm dark matter particles}

Depending on the WDM particle physics and properties, different scenarios can be considered regarding the particle velocity
dispersion as function of redshift.  In many scenarios, particles are copiously created in ultra-relativistic conditions and very
good thermal equilibrium.  As the Universe expands they may subsequently decouple on the thermal point of view because their
collisional relaxation rate becomes lower than the expansion rate, while still interacting with the rest of matter by gravitational
coupling.  If the particles do not decay their comoving number density is conserved, and if they follow the collisionless Boltzmann
equation their phase space density is conserved too.  But this later assumption is more fragile because some residual elastic
collisional relaxation processes can still decrease the effective phase space density by coarse graining.  

\subsubsection{Inspection of a commonly used result}
\label{sect:inspect}
A frequently cited derivation of WDM particle velocities as function of mass and redshift can be found in \cite{Bode01}.
In their Appendix A they recall that for a thermal relict particle X that decouples when relativistic, the abundance
$n_x$ relative to photons is:
\begin{equation}
\label{eq:abund}
\frac{n_X}{n_\gamma}=\left(\frac{43/4}{g_\mathrm{dec}}\right)\left(\frac{4}{11}\right)\frac{g_X}{2}
\end{equation}
where $g_\mathrm{dec}$ is the number of relativistic species present at decoupling, and $g_X$ is the number of spin states of the
particle.  Connecting then the particle mass density $\rho_X = m_Xn_X$ with the cosmological parameters $\Omega_X\equiv
\rho_x/\rho_c$ and $h$, where the critical universal density $\rho_c \equiv 3H^2/8\pi G$, and the Hubble constant $H \equiv 100 h
\,\mathrm{km \,s}^{-1} \,\mathrm{kpc}^{-1}$, they derive,
\begin{equation}
\label{eq:omega}
\Omega_{X}h^2\approx\frac{115}{g_\mathrm{dec}}\frac{g_{X}}{1.5}\frac{m_{X}}{\mathrm{keV}}\ .
\end{equation}
We confirm this equation when using $n_\gamma = 413\,\mathrm{cm}^{-3}$.  Then the authors proceed to derive a velocity formula. Since the
distribution function of fermions without chemical potential $\mu$ is proportional to $(\exp(\epsilon_x/kT_X)+1)^{-1}$, they point
out that if the particles decouple from photons when still relativistic $\epsilon_X = \left( p_X^2 c^2 + m^2c^4\right)^{1/2}-mc^2$
can be replaced by $p_X c $ where $p_X$ is the particle momentum.  To keep phase space density constant clearly in the relativistic
regime $p_X$ must stay proportional to $T_X$.  But obviously as the regime passes to non-relativistic this argument does not hold,
the exact formula valid at all $T_X$ is
\begin{equation}
\label{eq:pscaling}
p_X^2 \propto \left( kT_X \over c\right)^2 + 2kT_X\ .
\end{equation}
Therefore it seems incorrect to assume that $p_X$ is proportional to $T_X$ also at low $T_X$.  The exact scaling from
Eq.~(\ref{eq:pscaling}) becomes $p_X^2 \propto 2\,kT_X$, or $0.5\, m_X v_X^2 \propto kT_X$, that is, the kinetic energy
$\epsilon_X$, not the momentum, scales as temperature at all redshifts.

Another problem is the derived constant for velocity. They assume that the distribution function scales as the non-thermal
distribution $(\exp(v/v_0)+1)^{-1}$, and give without detail $v_0$ (in their Eq. (A3)), 
\begin{equation}
\label{eq:velocities}
\frac{v_0(z)}{1+z} = \mathbf{.012}\left(\frac{\Omega_{X}}{0.3}\right)^{\frac{1}{3}}\left(\frac{h}{0.65}\right)^{\frac{2}{3}} 
                    \left(\frac{1.5}{g_{X}}\right)^{\frac{1}{3}}\left(\frac{\mathrm{keV}}{m_{X}}\right)^{\frac{4}{3}}\kms
\end{equation}
where $z$ is the redshift.  But eliminating $\Omega_{X}h^2$ in this previous equation using Eq.~(\ref{eq:omega}) (their Eq.~(A2)), we
obtain for a $m_X=1$\,keV particle (rounding also to 2 significant digits),
\begin{equation}
\label{eq:final}
\frac{v_0(z)}{1+z}\approx{\mathbf{0.12}\left(\frac{1}{g_\mathrm{dec}}\right)^{1/3}\frac{\mathrm{keV}}{m_{X}}}\kms \ .
\end{equation}
Thus we find $g_\mathrm{dec} = \mathbf{1000}\, (g_X/1.5)^{1/3}$. This is too high a value for $g_\mathrm{dec}$ to be endorsed, as
mentioned by the authors, by large entropy producing processes.  Since the value for $g_\mathrm{dec}$ varies linearly with the mass
of the particle in the given cosmological model (Eq.\,\ref{eq:omega}), it allows the elimination of $g_\mathrm{dec}$ from the final
expression for velocity%
\footnote{The authors cite a value of 688 for the number of relativistic degrees of freedom at the time of decoupling of a 1 keV
  particle, while then using the value of 1000. In Viel et al. 2005 this latter value is used, although a rigorous calculation gives
  903 as the exact value.}.
This high value used for $g_\mathrm{dec}$ leads to a significant decrease in the particle velocities, as shown in Table 1.

In the minimal standard model the number of the full set of particles is $\sim 107$ while in the minimal supersymmetric
standard model, the value is increased to $\sim 229$ \citep{Pierpaoli}. Previous studies like \cite{Colombi}\footnote{Interestingly this is the paper cited by \cite{Bode01} as reference for production mechanisms of WDM and their
  relation to cosmology} use a value of $\sim 100$ for right-handed neutrinos decoupling before the electroweak phase
transition at very high temperatures, while \cite{Pierpaoli} assume a conservative reference value of 150 for both
gravitino and a standard warm dark matter candidate like the massive neutrino.

The lower value for the velocity adopted by \cite{Bode01} has been used in most WDM simulations thereafter.
This value for $g_\mathrm{dec}$, however, is valid for a 1\,keV particle only if we assume that dark matter is made entirely by these
type of particles, as shown in Eq.~(\ref{eq:omega}). For the cases in which a certain warm dark matter particle represents only a
fraction of the total dark matter content, this value is different and Eq.~(\ref{eq:velocities}) needs to be scaled
accordingly. This aspect has been overlooked in some simulation studies of mixed dark matter, providing misleading results as we
will show in Section 2.3.

The next line following \cite{Bode01} Eq.~(A3) states: ``The rms velocity is $3.571v_0$''.  Recalculating the rms
velocity of the adopted distribution function $f = (e^{v/v_0}+1)^{-1}$ we find a slightly larger factor: 
\begin{equation}
\langle v^2\rangle = {\int_0^\infty\! 4\pi v^4 f \,dv \over \int_0^\infty\! 4\pi v^2 f\,dv } = 15 {\zeta(5)\over\zeta(3)} 
   v_0^2 \approx (3.5\mathbf{9}714v_0)^2 \ , 
\end{equation}
where $\zeta$ is Riemann's function.  The slight discrepancy (the 9 digit) appears thus as a misprint.  

In Appendix A we find that the largest correction factor for the rms speed of a Fermi-Dirac distribution with respect to a
Maxwellian distribution is 1.07, not $\approx 3.6$ as stated in \cite{Maccio2012}. The difference comes entirely from
the very non-thermal distribution.

\subsubsection{Another scenario for quantum semi-degenerate  particles}
In the previous \cite{Bode01} scenario, WDM particles are treated as localized mass bullets following Boltzmann's equation.  However,
at creation time they are also assumed to be ultra-relativistic and in thermal equilibrium with radiation and the rest of matter,
typically following a Fermi-Dirac distribution since the known stable particles are fermions.  This entails that their quantum
nature does play a role at birth, they are at least semi-degenerate.  Phase space density is high enough for the distinction between
classical and quantum particles to play a role. The non-local Pauli principle applies then, each particle ``knows'' about the state
of each other.  Now if phase space density is approximately conserved then particles remain semi-degenerate at all times, which is
inconsistent with the usual assumptions applied at low redshifts that they behave as classical particles.

The known neutrinos offer a good example that particles are quantum objects instead of localized mass objects.  Real neutrinos are
in addition of being fermions also in a superposition of three mass states.  Since mass states propagate at different velocities,
with time relict neutrinos are actually in a superposition of entangled mass states increasingly spread apart.  How gravitational
interaction with matter structures can destroy the coherence of these entangled states is a question that will need to be addressed
in future works.

Here we develop a procedure to calculate precisely the particle velocity valid in all relativistic regimes for fermions or bosons. 
The full distribution Fermi-Dirac or Bose-Einstein distribution reads \citep[e.g.,][]{Padmanabhan} 
\begin{equation}
\label{eq:finit}
f(\vec p) =\frac{g}{(2\pi \hbar)^3}{ 1 \over \exp((\epsilon-\mu)/kT)\pm 1 } ,
\end{equation}
where $\vec p$ is the particle momentum, $g$ the spin-degeneracy factor (of order 1 or 2), $\mu$ the chemical potential,
$\epsilon=\sqrt{p^2c^2+m^2c^4}-mc^2$ the particle energy, $m$ the particle mass, and $T$ the particle temperature. 
The comoving number density is calibrated according to a neutrino-like scenario where the particles are once coupled to 
the photon background and in thermal contact, at a time where gravitational perturbations are still linear. 

First, the assumption that the chemical potential $\mu$ is constant and negligible is not necessarily valid for identical fermions
which are created in a half-degenerate state.  The Pauli principle has for effect that identical fermions, even with negligible
interaction (like the weak nuclear force for neutrinos), possess an effective \textit{exchange potential}, also sometimes called
\textit{exchange-correlation potential} \citep[e.g.,][]{Atkins}.  In quantum chemistry and Density Functional Theory (DFT) the
exchange potential is well known to be essential in the Hamiltonian describing electrons around a nucleus, or electrons in
materials, although the exact form in different contexts is sometimes not well known.  In the cosmological context the exchange
potential changes the chemical potential as the spatial density of identical fermions changes.  This effective repulsive interaction
for fermions makes the collisionless assumption of free fermions much less obvious.  In \cite{Pfenniger} the effective interaction
of free fermions was illustrated by solving exactly the time-dependent Schr\"odinger equation for two free but identical fermions in
3D space starting as Gaussian wave packets.  In the quantum regime (high phase space density) these wave packets effectively
interact and are scattered due to the repulsive exchange potential.  In the classical regime (low phase space density) the
wave-packets follow, as expected, a straight trajectory.

In quantum statistical mechanics \citep[]{Huang} the exchange potential between two particles has a well known form
dependent on temperature and distance $r$
\begin{eqnarray}
\phi(r) &=&  -kT \log\left(1 \mp \exp\left(-mkT {r^2 \over \hbar^2}\right)\right)\\
     &=& -kT \log\left(1 \mp \exp\left(-2\pi {r^2\over \lambda^2}\right)\right),
\end{eqnarray}
where the $-$ sign applies for fermions and $+$ for bosons, and $\lambda$ is de Broglie wavelength.  For semi-degenerate particles
$\lambda$ is of order of $n^{-1/3}$, so in a semi-classical description fermions ``feel'' a rapidly varying repulsive force from
neighbouring particles, while bosons an attractive force.  The reality of the exchange potential can be invoked to cast a doubt that
the commonly assumed collisionless approximation for semi-degenerate particles is valid in the cosmological context.  Instead one
should expect a local thermalization of identical particles on a short time-scale.

To calculate the chemical potential evolution in the cosmological context, one needs therefore an additional assumption, besides
number conservation.  The particle momentum and kinetic energy can not be assumed conserved due to the global gravitational
interaction.  A reasonable assumption \citep[]{Trodden} is that the expanding
medium proceeds adiabatically, at least as long as gravitational clustering is linear.  This means that the particle specific
entropy can be taken as a conserved quantity.

The entropy $S$ expressed as a function of other thermodynamical variables reads \citep[e.g.,][Vol. I, Eq. 5.73]{Padmanabhan},
\begin{equation}
\label{eq:entropy}
S= {1 \over T}\left(E+PV-\mu N \right) \ ,
\end{equation}
where $E$ is the total thermal energy, $P$ the pressure, $V$ the volume, $\mu$ the chemical potential, and $N$ the number of
particles.  The specific entropy $s\equiv S / N$ divided by Boltzmann's constant $k$ is a pure number
\begin{equation}
\label{entropy2}
{s(T,\mu) \over k} ={1 \over kT}\left({e+P\over n} -\mu \right)
\end{equation}
where $e=E/V$ is the specific energy density and $n=N/V$ is the number density.

The thermodynamical quantities for fermions and bosons at all regimes can be calculated accurately by evaluating numerically the
relativistic Fermi-Dirac or Bose-Einstein integrals for particle density $n$, energy density $e$, and pressure $P$
\citep[e.g.,][Vol. II, p. 216]{Padmanabhan} as functions of temperature $T$ and chemical potential $\mu$
\footnote{This part follows closely the calculations made in \cite{Pfenniger}, but correct a mistake where the used entropy
  expression was only valid in the ultra-relativistic regime, or when $\mu=0$.}
:
\begin{equation}
n(T,\mu) = {4\pi g\over h^3} \int_0^\infty {p^2 \over \exp((\epsilon -\mu)/kT) \pm 1}\, dp\ ,
\end{equation}
\begin{equation}
e(T,\mu) = {4\pi g\over h^3} \int_0^\infty {p^2\epsilon  \over \exp((\epsilon -\mu)/kT) \pm 1}\, dp\ ,
\end{equation}
\begin{equation}
P(T,\mu) = {4\pi g\over h^3} \int_0^\infty {p^2 \over \exp((\epsilon -\mu)/kT) \pm 1} {1 \over 3} {c^2p^2\over \epsilon+mc^2}\, dp\ ,
\end{equation}
where $g$ is the number of distinct particle states, and $\epsilon = \sqrt{p^2 c^2 + m^2 c^4} - mc^2$ the particle energy.
In the integrands the $+$ sign is for fermions, the $-$ sign for bosons.
The conserved particle density $n(T,\mu)$ is related to universal expansion by the scale factor $a = 1/(1+z)$, thus 
\begin{equation}
  \label{nz}
  n(T(z),\mu(z)) = n_0 (1+z)^3 \ ,  
\end{equation}
while the constant particle entropy gives 
\begin{equation}
  \label{entropyz}
  {s(T(z),\mu(z))\over k} = {s(\infty,0)\over k} = 4.20183245 \ ,  
\end{equation}

For Fermi-Dirac particles the solution of this system for $n_0=115 \,\mathrm{cm}^{-3}$, $m= 1\,$keV, $g=1$ in the non-relativistic
regime is :
\begin{equation}
  \label{solFD}
   {\mu\over kT} =  -1.6202, \quad {mc^2\over kT} = 5.6186 \cdot 10^{12}\ .
\end{equation}
For a graphical illustration of these functions behaviour, see Fig.\,\ref{fig:FCTS} in Appendix B.
For a couple of dex around this solution the scaling with $n$, $g$ and $m$ for $T$ and $v$ goes with good approximation as follow
\footnote{An astute reader might notice that for classical massive neutrinos ($0.01 < mc^2/\mathrm{eV} < 2$) the found temperature
  is much lower than the commonly quoted temperature of 1.9\,K.  Actually the 1.9\,K value is valid for massless neutrinos only.
  The difference comes from the misleading use of temperature as an equivalent concept for energy and vice versa, while the
  neutrino rest mass energy does not contribute to thermal energy. The proper meaning of temperature is the quantity that would be
  measured, in the case of real neutrinos, by a cosmic sized thermometer able to thermalize with the neutrino background. }
:
\begin{eqnarray}
  \nonumber 
  T &=&  2.0654\cdot 10^{-6}\left({n \over 115\,\mathrm{cm}^{-3}}{1 \over g}\right)^{2/3} \left(mc^2\over \mathrm{keV}\right)^{-1} \,\mathrm{K}\, \\
\label{eq:twentysix}
  {v} &=&  0.2226\, \left({n \over 115\,\mathrm{cm}^{-3}}{1 \over g}\right)^{1/3} \left(mc^2\over \mathrm{keV}\right)^{-1} \kms \ .
\end{eqnarray}
When the regime becomes relativistic this approximation is no longer accurate.  One can solve the pair of equations
(\ref{nz}) and (\ref{entropyz}) in any situation.

In comparison, for Bose-Einstein particles the solution for the same parameters is:
\begin{equation}
  \label{solBE}
    {\mu\over kT} =  -1.2451, \quad  {mc^2\over kT} = 8.1348 \cdot 10^{12} \ .
\end{equation}
Around this solution the scaling with $n$, $g$ and $m$ for $T$ and $v$ goes approximately as:
\begin{eqnarray}    
  \nonumber 
  T &=&  1.4265\cdot 10^{-6} \left({n \over 115\,\mathrm{cm}^{-3}}{1 \over g}\right)^{2/3}  \left(mc^2\over \mathrm{keV}\right)^{-1} \,\mathrm{K} \, \\
   {v} &=& 0.1768\, \left({n \over 115\,\mathrm{cm}^{-3}}{1 \over g}\right)^{1/3} \left(mc^2\over \mathrm{keV}\right)^{-1} \kms \ .
\end{eqnarray}
If Maxwell-Bolzmann particles are used in simulations one can also calculate the solution, replacing the $\pm1$ in integrals by
zero, and taking $s(\infty,0)/k = 4$. The velocity coefficient is found to be $0.20592 \kms$, intermediate  between the Fermi-Dirac and
Bose-Einstein cases.  The correction of quantum statistics with respect to a Maxwellian distribution remains thus small, as demonstrated in Appendix A.



\subsection{Power spectrum of the warm dark matter simulations}

Since collisionless physics does not depend on the particle mass, the power spectrum must directly depend only on the velocity
distribution of the particles, which results from the particle production mechanism. \cite{Colombi} and \cite{Bode01} also emphasize
this point.  

To compute the transfer function for WDM models the fitting formula suggested by Bode, Turok and Ostriker (2001) gives:
\begin{equation}   
T^2(k) = {P^{\mathrm{WDM}} \over P^{\mathrm{CDM}}} = [1+(\alpha k )^{2 \nu}]^{-10/\nu}   
\label{eq:Twdm}   
\end{equation}   
where $\alpha$, the scale of the break, is a function of the WDM parameters, which are function of the velocity, while the index
$\nu$ is fixed. People prefer however, to use the mass dependence instead of the velocity, using Eq.\ (\ref{eq:final}) as a
conversion. 

\cite{viel05} (see also \cite{hansen02}), using a Boltzmann code simulation, found that $\nu=1.12$ is the best fit for $k<5 ~h~ \rm Mpc^{-1}$, and
they obtained the following expression for $\alpha$:
\begin{equation}   
\alpha = 0.049  \left ( {m_x \over  \rm{1\,\mathrm{keV}}} \right )^{-1.11}   \left ( { \Omega_{\nu}  
\over 0.25 }\right )^{0.11}   \left ( { h \over 0.7} \right )^{1.22} \mpch.   
\label{eq:alpha}
\end{equation}   

In the case of warm dark matter particles, the streaming velocity supresses the matter power spectrum $P(k)$ and the formation of
structure, on scales smaller than their free-streaming scale. A rough estimation of the free-streaming scale is given by \cite{Bond}:
\begin{equation}
\label{scale}
k_{\rm FS} = {2\pi \over \lambda_{\rm FS}} \sim
5 \, \mathrm{Mpc}^{-1} 
\left({m_{x} \over 1 \, \mathrm{keV}}\right)
\left({T_{\nu} \over T_{x}}\right)~,
\end{equation}  
Depending on the model for the properties of a certain particle, there can be different expressions for the damping of the power
spectrum \citep[e.g.]{aba06}, but for the purpose of our present work and for easier comparison with previous studies we use the
expression given in Eq.\ (\ref{eq:alpha}) with the corresponding thermal velocities.

This approach used for cutting the power spectrum is only valid however, for a scenario in which the whole dark matter content is
made up by one specific dark matter particle of a certain velocity.

\subsection{Caveat Emptor}

In this section we would like to summarize the findings of previous sections and discuss some of their implications.  We want to
stress that the assumed particle production model and physics strongly impact on the ascribed particle mass, while the initial
velocity distribution and its corresponding power spectrum is the only really important initial parameter influencing the simulation
results.  As far as the physics behind the origins of the dark matter particles is concerned, the assumptions found in the
literature can widely differ.

Previously on section 2.2.1. we showed how the \cite{Bode01} result for estimating velocities for neutrino like particles is based
on arguments like entropy production and negligible chemical potential.  The expression for velocities in Eq.\ (\ref{eq:velocities})
is based on a dependence of the number of species on the mass of the particle, such as to preserve the equivalence in Eq.\
(\ref{eq:omega}).  For the models with cold plus warm dark matter, or models with different particle mass, the value of the number
of species should be adjusted accordingly. Many papers that study mixed particles simulations have omitted this readjustment for
velocities \citep[e.g.][]{Ander}. Eq.\ (\ref{eq:velocities}) has been reduced by the fraction with which a certain particle
contributes to the total density, therefore leading to inconsistencies like having for a certain mass of a particle, different
thermal velocities, depending solely on that fraction.  Moreover, since the power spectrum cutoff depends on the velocity of the
particle (not the mass), studies that use the cutoff for a velocity, but a different thermal velocity, given by a different model of
particle production are not consistent.  These results, although used for constraining the mass of a particle in terms of detection
experiments, should not be considered as accurate.

\begin{table}
  \caption{Correspondence between particle mass $m$ and rms velocity dispersion in literature for 0.2, 1 and 3.5 keV. 
    The first column shows the value originally given in Bode et al.\ (2001), the second column shows the value obtained 
    using $g_\mathrm{dec}=150$ \citep{Pierpaoli} in Eq.\ (\ref{eq:velocities}), and the third one, the value given by our derivation.
  }
\label{table:velev}
\begin{center}
\begin{tabular}{@{}l|ccc}
\hline
\hline
Mass & Bode et al. & Pierpaoli et al. & This work\\ 
     & Eq. (11)$\times 3.571$ &  & Eq.~(25)\\ 
\hline
keV$/c^2$ & km/s & km/s & km/s\\
\hline
0.2 & 0.366 & 0.4032 & 1.113 \\
1.0 & 0.0429 & 0.0225 & 0.223 \\
3.5 & 0.00806 & 0.0230 & 0.0636\\
\hline
\end{tabular}
\end{center}
\end{table}

As an alternative, we provide a different energy-thermal velocity correspondence based on number conservation and a non-entropy
production while taking into account the quantum pressure, but assuming a thermalization caused by the exchange potential.  Entropy
conservation by particles in the hot Big Bang is invoked by many authors, such as \cite{Padmanabhan} or \cite{Weinberg08}.
From Eq.\ (\ref{eq:twentysix}) which estimates the thermal speed of WDM particles, independent of the cosmological parameters, we
have the following velocity dependence with the redshift:
\begin{equation}
  \label{solFD2}
  {v\over 1+z} =  \mathbf{0.2226}\, \left({n \over 115\,\mathrm{cm}^{-3}}{1\over g} \right)^{1/3} \left(mc^2\over \mathrm{keV}\right)^{-1} \kms \ .
\end{equation}
The difference between our and \cite{Bode01} estimations is showed in Table.\ (\ref{table:velev}) for 0.2, 1 and 3.5 keV
respectively, at redshift zero.  The \cite{Bode01} speed for 1\,keV fermions out of equilibrium, used in most WDM simulations, is 5
times lower than the value derived here. In general these differences cumulate their effect if simulations are started at higher
redshifts, and are crucial not only for phase space density studies, but also for structure formation.

Our finding affects the results and conclusions of previous papers which were using Eq.\ (\ref{eq:velocities}) to constrain the mass
of sterile neutrinos. This extends even to papers which did not include thermal velocities.  The power spectrum studies based on the
velocity of the particle are subject to the same difference in the velocity estimation (see Section 2.3).  Also, when comparing the
thermal velocities to the Zeldovich velocities, these factors weaken correspondingly the reason for ignoring the thermal velocities,
against all the arguments presented in Section 2.1.

Since our aim here is to describe typical qualitative effects on structure formation present in a broader range of energies, we will
refer to the particles in terms of their velocity dispersion instead of their mass.  Indeed the thermal velocity of a particle as
its decoupling temperature at a certain redshift depends on the specific physics of particle production. That influences the
ascribed mass of that particle. More complex analysis of the decoupling theories for a certain particle may give a slightly
different dependence between the thermal velocity at a certain redshift and the particle mass.

\begin{figure*}
\begin{tabular}{ccc}
\psfig{file=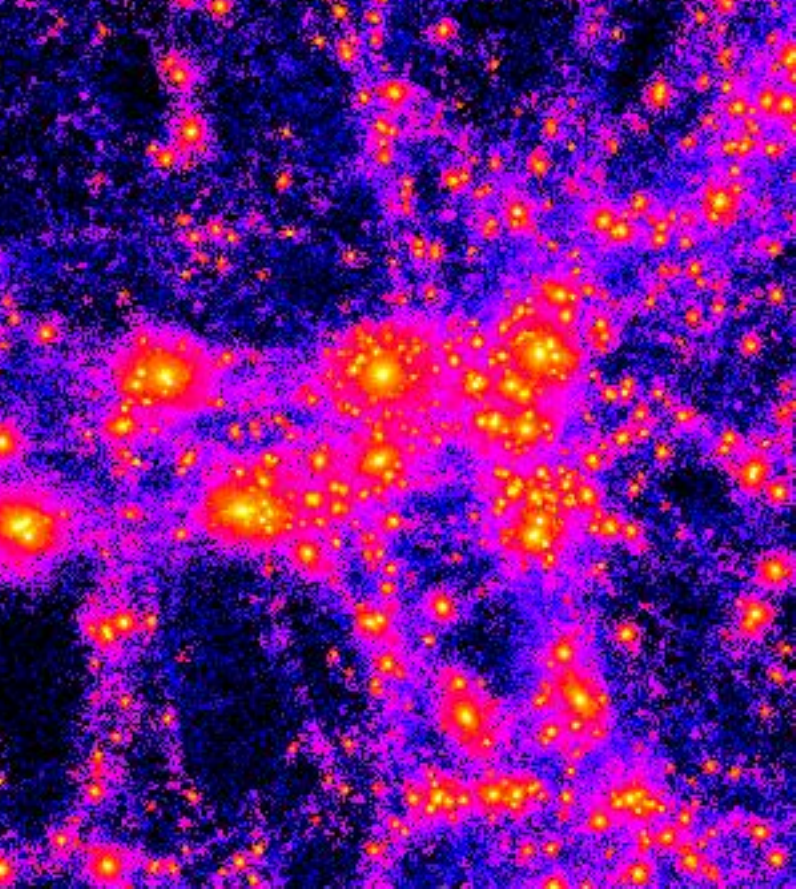,width=160pt,height=160pt} &
\psfig{file=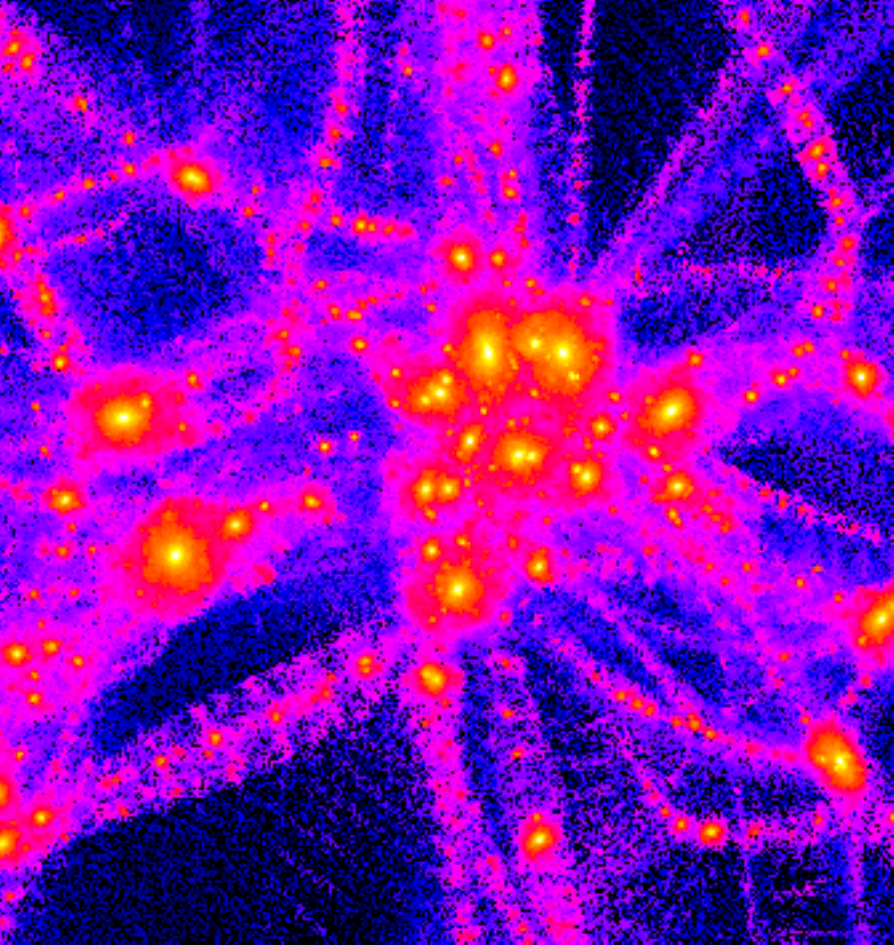,width=160pt,height=160pt} &
\psfig{file=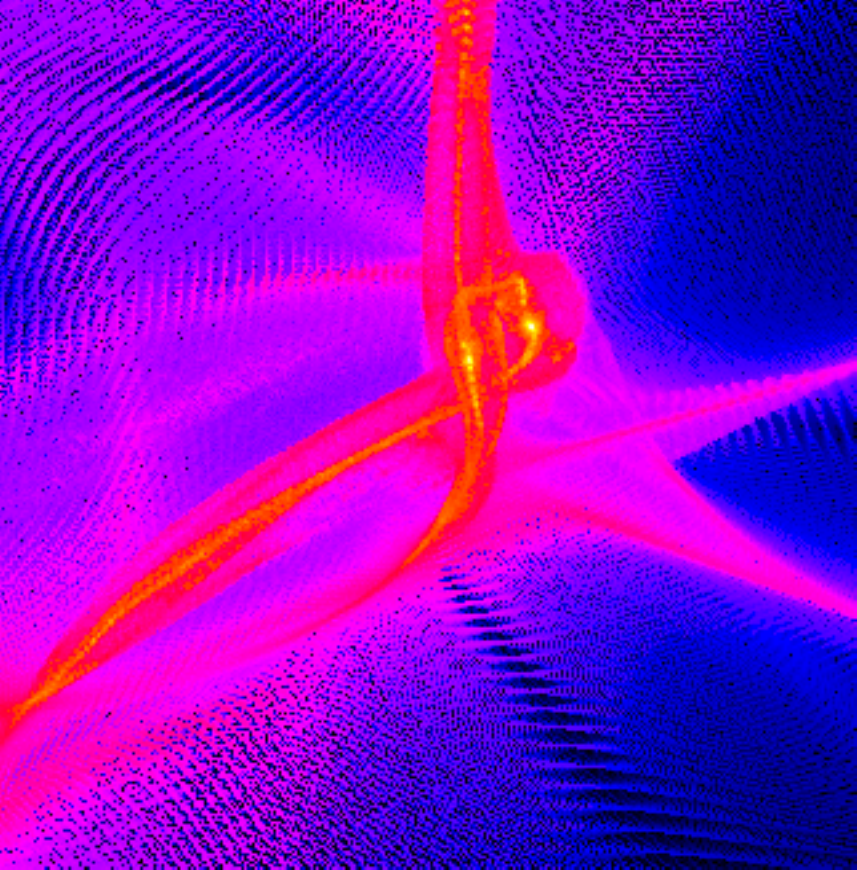,width=160pt,height=160pt} \\ 
\end{tabular}
\caption{Illustrative density map of structure formation regions at redshift zero, from left to right, in CDM, WDM (0.3\,km/s) and HDM (2.3\,km/s). For a similar illustration of a full box see Macci{\`o} et~al (2012). These simulations have not been used in this paper.}
\label{fig:dm}
\end{figure*}

\section{Simulations setup}

We conducted several suites of $N$-body simulations.  All simulations have been performed once with {\sc pkdgrav-2}, a treecode
written by Joachim Stadel and Thomas Quinn \citep{stadel01}\footnote{http://hpcforge.org/projects/pkdgrav2/} and then using Volker
Springel's Gadget-2 \citep{springel05}\footnote{http://www.mpa-garching.mpg.de/gadget/}.  The initial conditions are generated with
Ed Bertschinger's {\sc grafic2} package \citep{bertschinger01}\footnote{http://web.mit.edu/\~{}edbert/}. Although some differences
have been spotted between the two different codes, those differences are not qualitatively important where structure formation is
concerned, there can be minimally spotted at very small scales.

The simulations we have performed cover a range of velocities from 0.01\,km/s to 10\,km/s (3.5\,keV to 15\,eV) at
redshift zero. For illustration purposes, in Fig.~\ref{fig:dm} we show generic density maps of structure formation regions in CDM, WDM and HDM simulations.
 Particles that have $\sim 0.1\,\mathrm{km/s}$ velocity dispersion are in a transient regime from a
predominant top-down structure formation scenario to a hierarchical one, showing both these trends. We have chosen one
such simulation and compared it to a simulation of a colder particle more favored by the observational constraints and
with a cold dark matter simulation. For the warm dark matter the simulations have been prepared with initial power
spectrum consistent with initial velocities, and, for comparison, the same initial power spectrum without initial
velocities.

\begin{figure*}
\begin{tabular}{cc}
\psfig{file=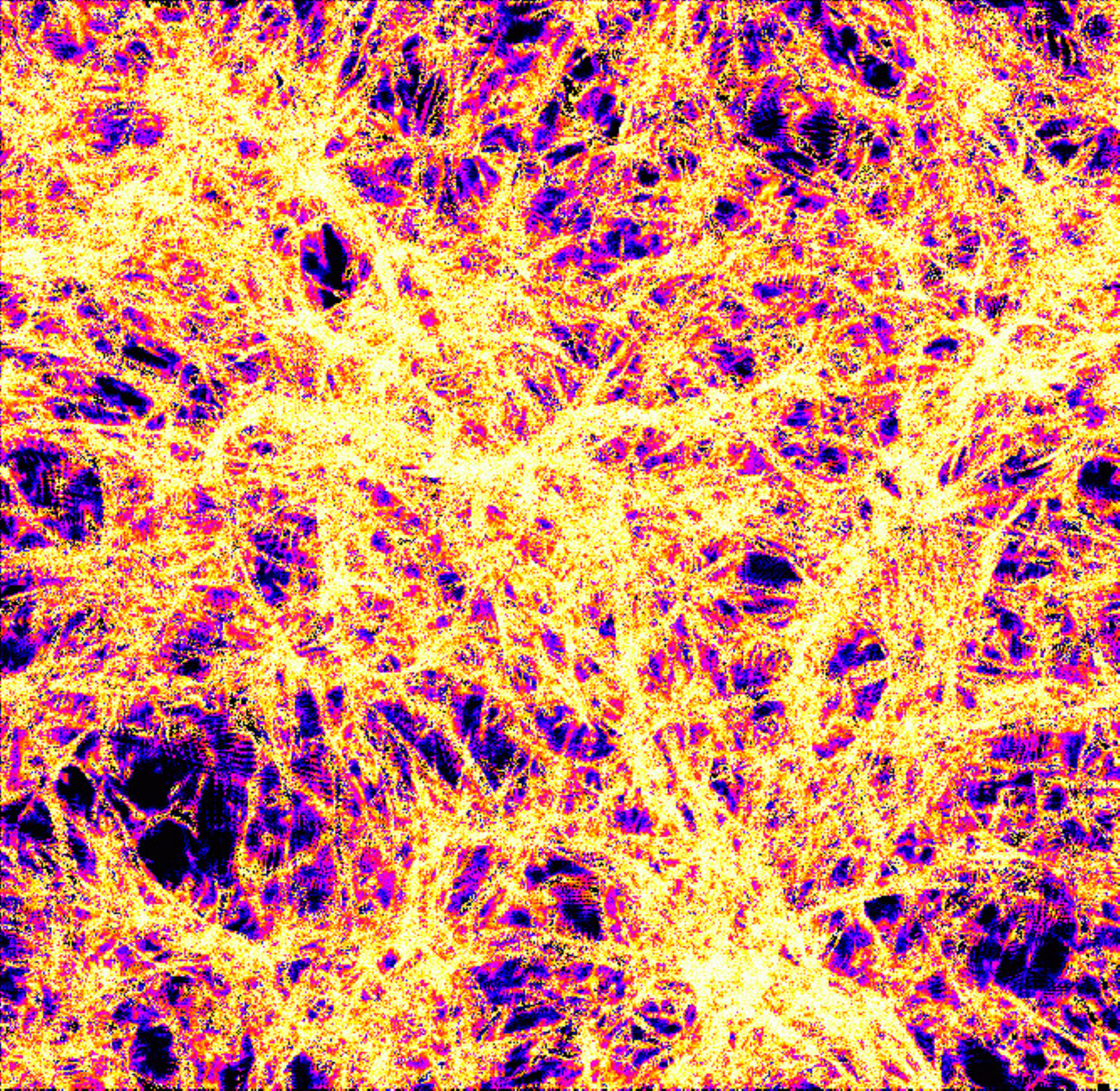, height=200pt} &
\psfig{file=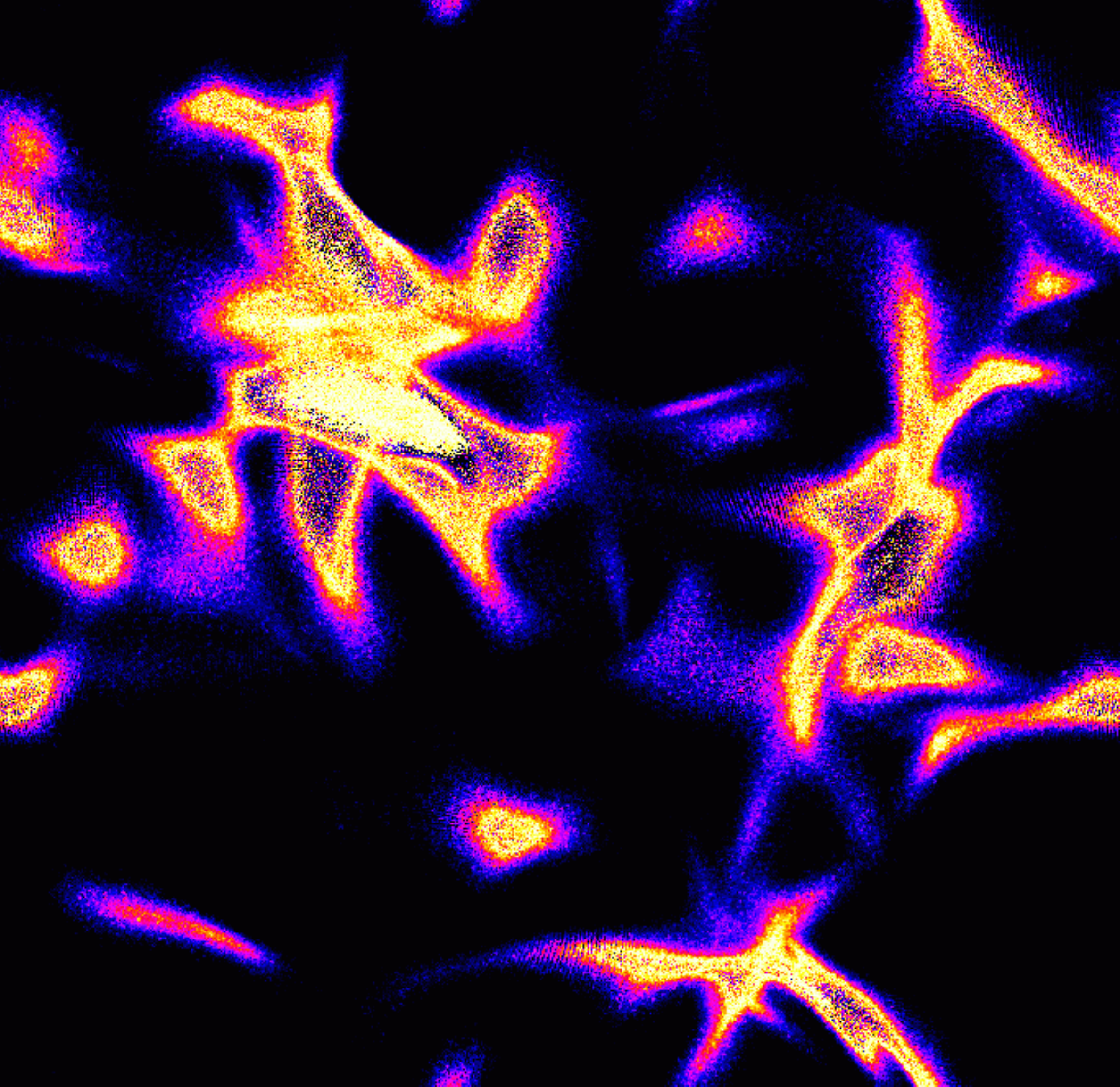, height=200pt} \\
\end{tabular}
\caption{Plot of high density regions at redshift 2.3 in a 40\,Mpc/h simulation box with CDM particles in the left panel and WDM
  particles (model WDM2) in the right one, showing major topological differences}
\label{fig:clip}
\end{figure*}

\begin{table}\caption{Details of the simulations}
\label{table:sim}
\begin{center}
\begin{tabular}{@{}lccccc}
\hline
\hline
Label & velocities $z_i$ & cutoff & box size & N & softening\\ 
 & \,km/s & \,eV & \,\Mpc/h & & pc \\
\hline
CDM & no & - & 40 & $300^3$ & 50 \\
\hline
WDM1 & no & 200 & 40  & $300^3$ & 50 \\
WDM2 & 36.6 & 200 & 40 & $300^3$ & 50 \\  
WDM3 & no & 1000   & 40 & $300^3$   & 50 \\
WDM4 & 4.6 & 1000 & 40 & $300^3$ & 50  \\
\hline
WDM5 & 36.6 & 200 & 30 & $256^3$ & 100 \\
\hline
\hline
\end{tabular}
\end{center}
\end{table}

The simulation parameters are summarised in Table \ref{table:sim}. The starting redshift for the simulations is $z_i=100$ in order
to ensure a proper treatment of the non-linear growth of cosmic structures.

The cosmological parameters used are given by the WMAP7 results: $\Omega_{\Lambda}$=0.721,
$\Omega_m$=0.279, $\Omega_b$=0.0463, $h=0.7$ and $\sigma_8=0.821$, 

We start with running large scale simulations in cosmological box of 40\,Mpc/h, engaging $300^3$ dark matter particles
and one 30\,Mpc/h box with $256^3$ particles.  We then select a region where the top-down halo formation is predominant
and re-simulate it with an eight times higher resolution.

\section{Simulations analysis and results}

\subsection{Structure formation in WDM simulations}

Free streaming causes a delay in the formation of structure in the warm dark matter simulations. This delay depends on the energy,
hence the velocity of the particle. The higher the thermal velocity of the particle, the later the filaments will reach the
collapse, making it impossible for structures to be formed by redshift zero in hot dark matter scenarios, as illustrated in
Fig.~\ref{fig:dm}, right panel.

The fragmentation of the filaments is observed in all N-body simulations of warm dark matter where the collapse is stimulated by the
noise in the particle distribution \citep{Bode01, gotz, Wang}. This is especially observable at the characteristic grid or glass
spacing. Above the free streaming scale the mass function is flattened to a value that closely matches the luminosity function of
galaxies (assuming mass traces light). The length and the lifetime of the filaments depend on the energy of the particle. The
higher the velocity dispersion of the particle, the larger will be the filaments and the longer they will be preserved. These can reach
20\,Mpc in a 40\,Mpc box and survive until redshift zero in the case of velocities of few km/s and above.

In Fig.\,\ref{fig:clip} the difference between high density regions in a CDM simulation, versus WDM at redshift 2.3 is shown.  The
picture displays the 2D projection of the 3D density map of the full simulation box. One can see that due to the free streaming,
particles are concentrated in large spatial structures, delimited from each other by large low density regions, or voids, as opposed
to the crowded web present in the cold dark matter simulation.

It is well known that in CDM models, smaller halos collapse first and merge hierarchically into larger systems, as it is obvious in
all high resolution simulations \citep{DiemandMoore, diemandstructure, stadel09}. Furthermore, one finds that less massive halos
are more concentrated, perhaps reflecting the fact that the density of the universe is higher at earlier epochs, since the CDM
particles have an infinite phase space density.

On the other end of the velocities spectrum, for HDM models, the structure formation is essentially top-down up to redshift zero, as can be seen in the right panel of Fig.~\ref{fig:dm}, with just large filaments collapsing into few halos.

As stated in the introduction, in the case of warm dark matter we see from our simulations that the structure formation is more
complex, a hybrid mechanism where both long-range and short-range effects are present, from long distance to nearest neighbours,
from top-down to bottom-up.

The top-down trend predominant in the early epochs in warm dark matter simulations has been missed in previous works since it is
difficult to observe it while analyzing the snapshots of the simulations.  For particles with velocities smaller than a few km/s the
top-down trend is hidden by the hierarchical growth that dominates at later times. We have been able to see this effect in our
simulations while watching movies made with a sufficient large number of snapshots.  We stress that only movies convey the
complexity of these multiscale hierarchical processes.  Several movies of WDM structure formation, filament collapse and halo formation from this study can be found on a youtube playlist (\url{https://www.youtube.com/playlist?list=PLnGS4wkStJ1aqi3M9hTDaUzuZ-vs-Qg6i})\footnote{All the HD movies are on youtube and can be watched indivadually on this channel \url{https://www.youtube.com/channel/UCEmQi8hDNW2emqGn-urtvpg}. Remember to adjust your settings to HD quality. Links for direct download can be provided on demand. For a short description of the movies and movies snapshots, see Appendix C.}.

As the movies show, in WDM simulations, structures form in a qualitatively different way from CDM models.  The hybrid structure
formation is more complex than what the traditional top-down/bottom-up dichotomy can categorize, as discussed in the Introduction.

\begin{itemize}
\item{During the early stages one sees the formation of well contoured filaments. How early is this stage depends on the particle
    velocity. In our WDM2 simulations this happens in the interval $13> z >8$.}
\item{In the higher density regions, usually situated at the intersection of such filaments, the first halos are formed through
    gravitational collapse. These halos continue growing into larger ones by accreting particles from the disrupted filaments
    (Fig.\,\ref{fig:C1}).}
\item{In medium density regions, halos show a hierarchical formation trend. Small halos collapse first and then merge into bigger
    halos (Fig.\,\ref{fig:C2}).}
\item{In less dense regions, the ones isolated by voids and which have a very slow evolution, we have observed filaments that
    collapse very late. The top down formed halo survives without any mergers until redshift zero (Fig.\,\ref{fig:C3}).}
\item{Finally there is the more complex scenario in which we observe large halos formed earlier which merge together forming a
    large cluster (Fig.\,\ref{fig:C4}).}
\item{The filamentary-like structure is preserved until redshift zero, with new filaments forming in the low density regions as late
    as redshift $z\sim4$.}
\end{itemize}

Looking closer, we have analyzed four different regions in our simulations and displayed them in four different movies. The
characteristics of these regions are summarized in Table \ref{table:mov}.

\begin{table}
\caption{The properties of four different regions of simulation WDM2 displayed in the movies}
\label{table:mov}
\begin{center}
\begin{tabular}{@{}lcccc}
\hline
\hline
Label & size & first collapse & average density & highest density\\ 
\hline
 - & box & z & critical & critical \\
\hline
lu.avi & 1/4 & 10.13 & 0.264 & 477 \\
ld.avi & 1/4 & 9.45 & 0.258 & 481 \\                                                                
ru.avi & 1/4& 10.77 & 0.268 & 480 \\
rd.avi & 1/4& 9.78 & 0.258 & 474 \\
\hline
\hline
\end{tabular}
\end{center}
\end{table}

\begin{figure*}
\begin{tabular}{cc}
\psfig{file=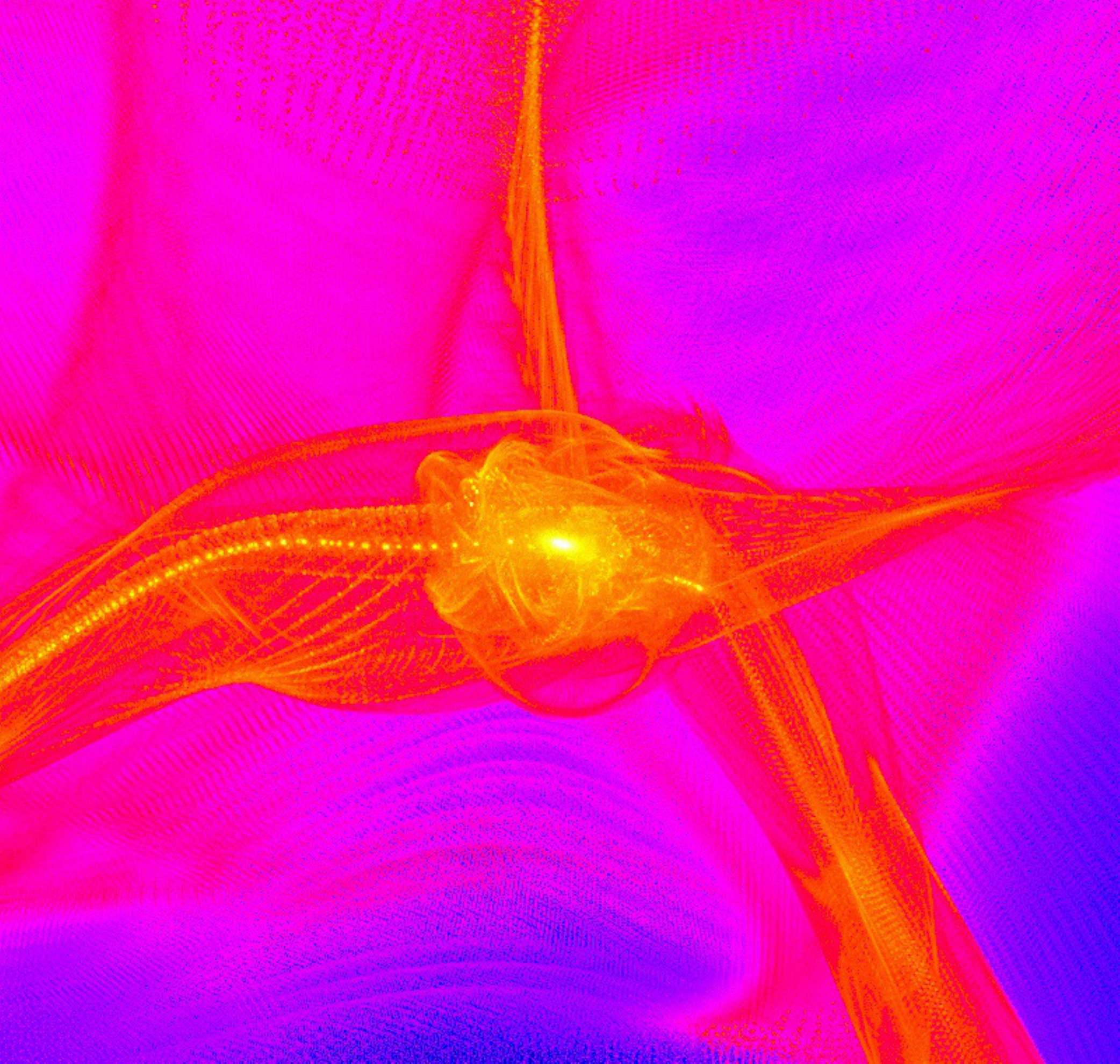, width=200pt, height=200pt} &
\psfig{file=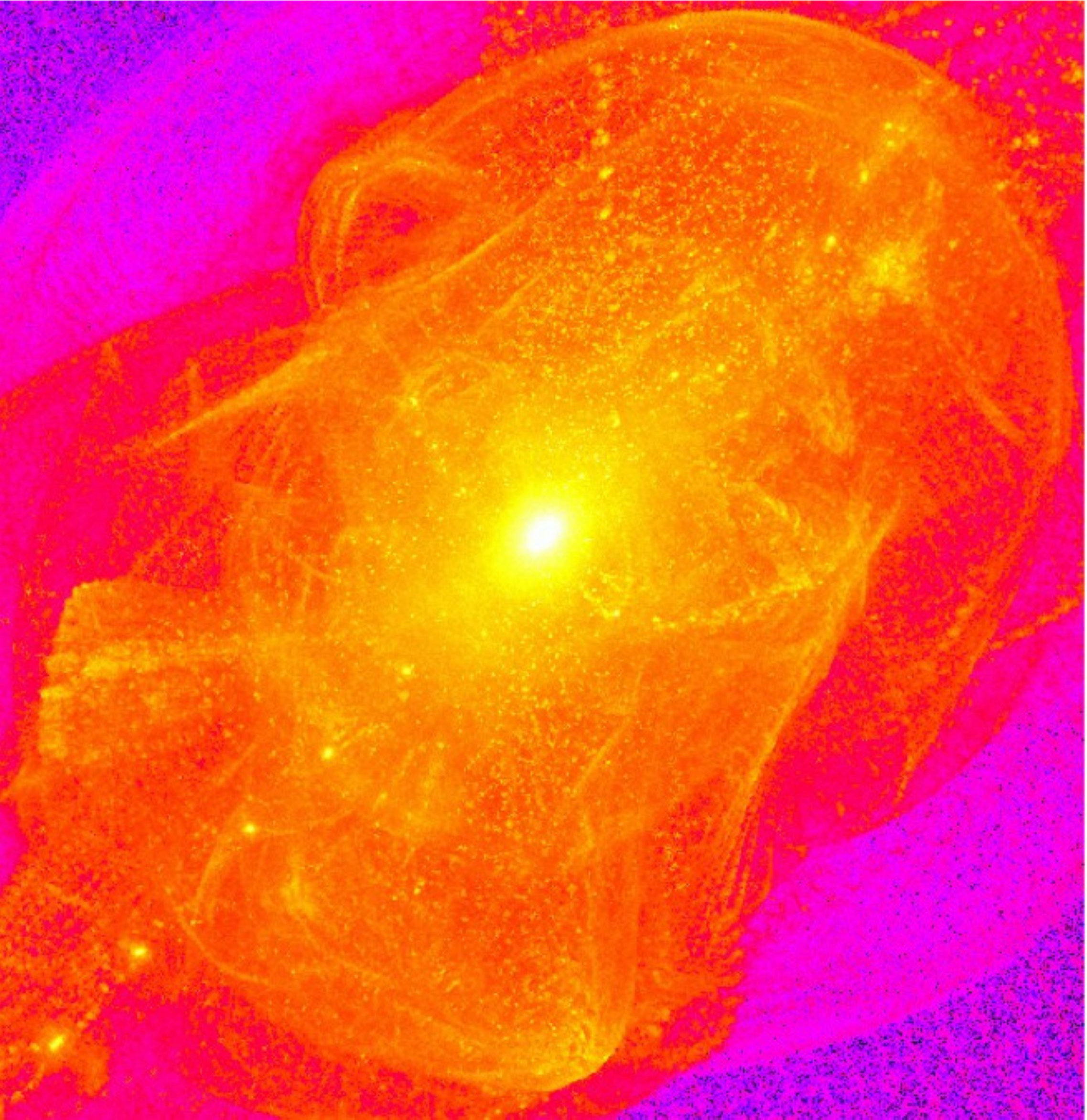, width=200pt, height=200pt} \\
\end{tabular}
\caption{Halo formation at the intersection of filaments. A zoom in projection shows that shells and caustics are visible in the not yet virialized WDM halo.}
\label{fig:halotop}
\end{figure*}

\begin{figure*}
\begin{tabular}{cc}
\psfig{file=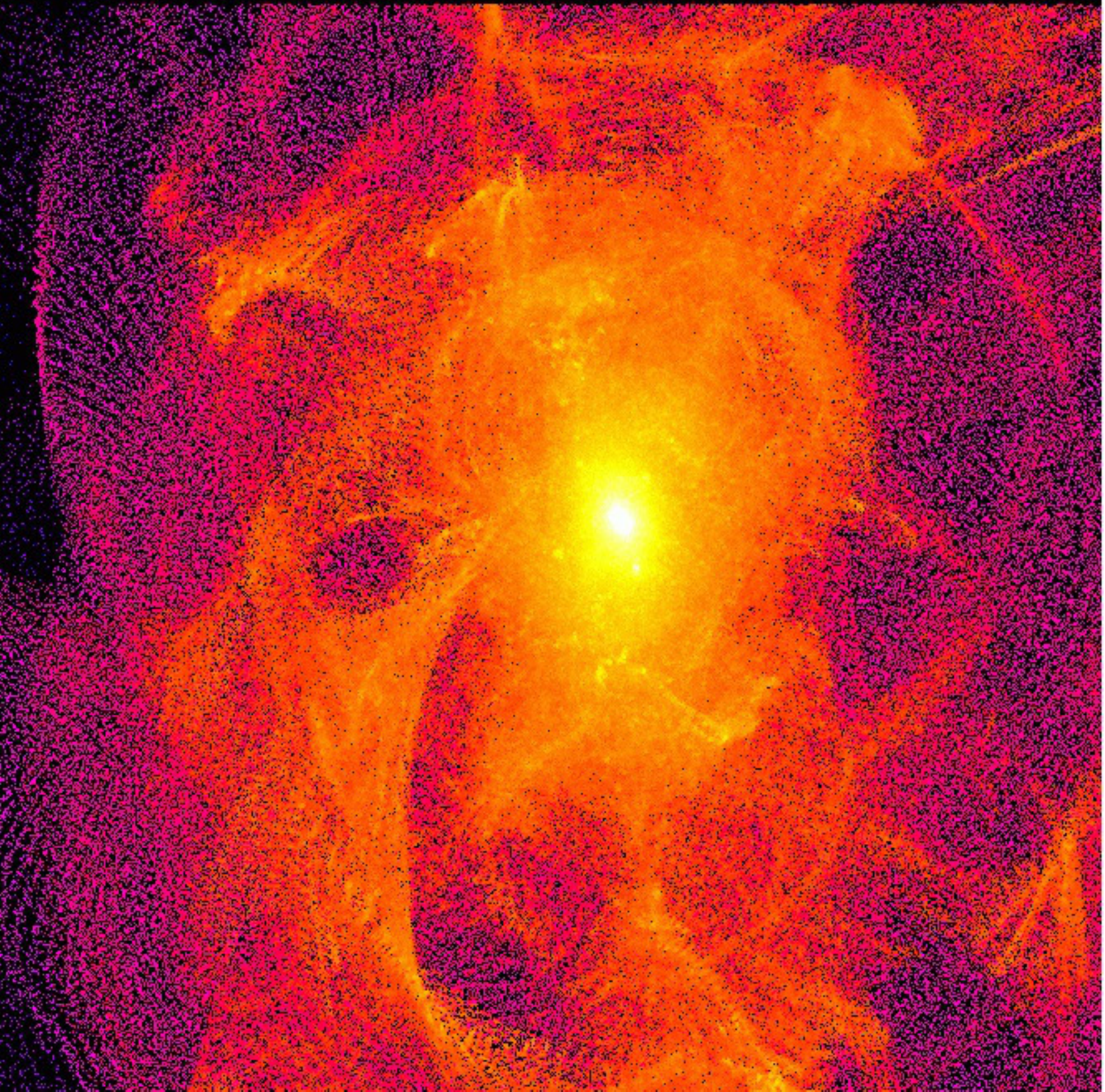, width=200pt, height=200pt} &
\psfig{file=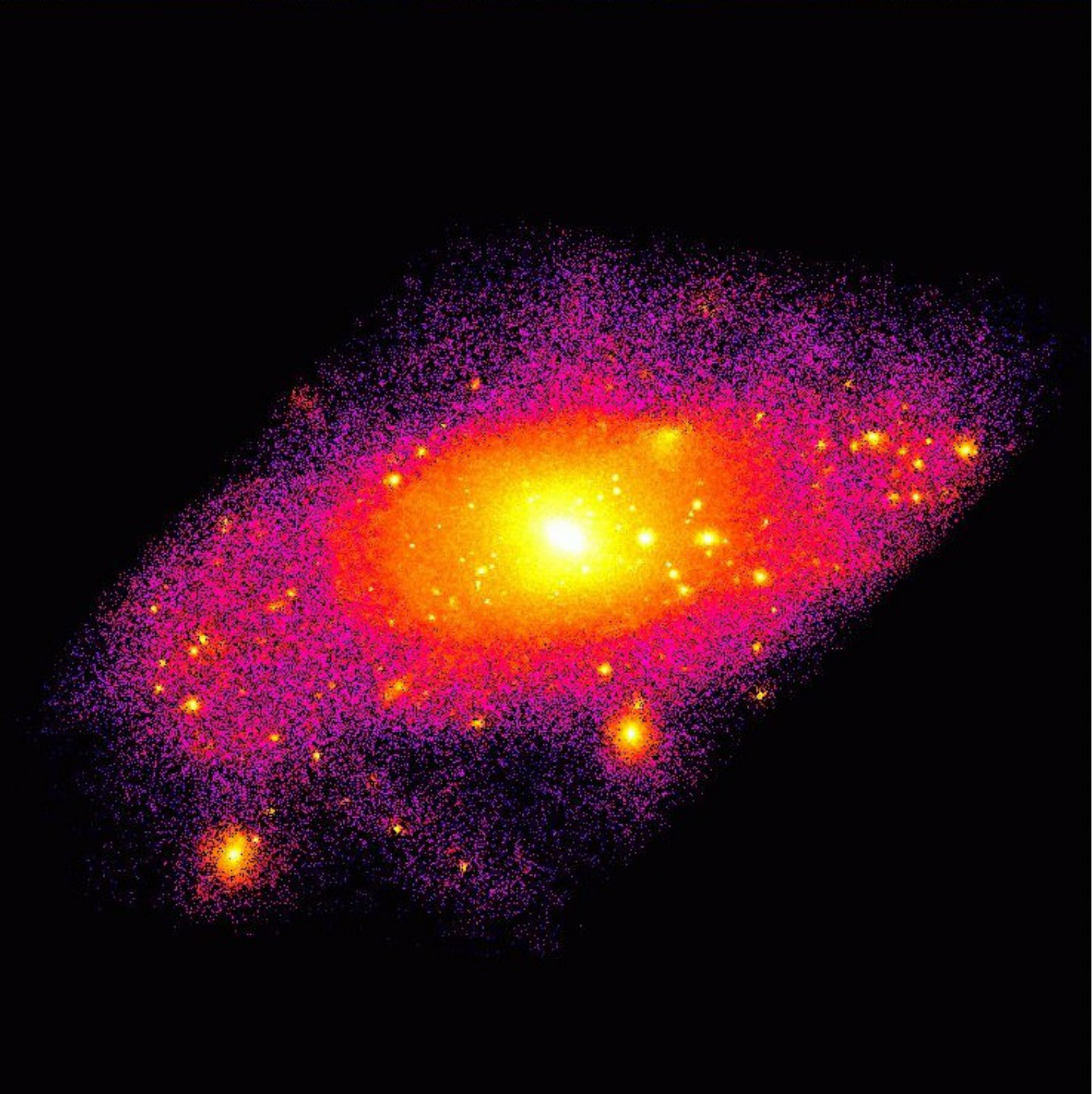, width=200pt, height=200pt} \\
\end{tabular}
\caption{A thin slice of the WDM halo formed Top-Down on the left and of a CDM one on the right. Very different internal structure, with shells in caustics in the WDM halo being more apparent}.
\label{fig:slice}
\end{figure*}

Our conclusion from analyzing these simulations is that there is only one correlation, between the time of the first collapse and
the density reached in a certain region, and that depends only on the network morphology and architecture of that region. The first
halos collapse in the region where the density becomes $\sim 2 \times 10^3$ times larger that the average density and almost
$\sim 3\times 10^3$ times larger than the lowest densities present in that region at that epoch. In the simulations with particle
velocity of 0.36 km/s (that mimic 200\,eV), the first collapse appears just after redshift 10 with the first halos forming until
redshift 4, while in the simulations with 0.04 km/s (that mimic 1\,keV), the first structures would have been already formed by
redshift 10. The first halos form at the high density region at the intersection of the filaments and then continue accreting
matter. If in a certain region there are many filaments collapsing, then the halos will merge into bigger ones.

\begin{figure*}
\begin{tabular}{cc}
\psfig{file=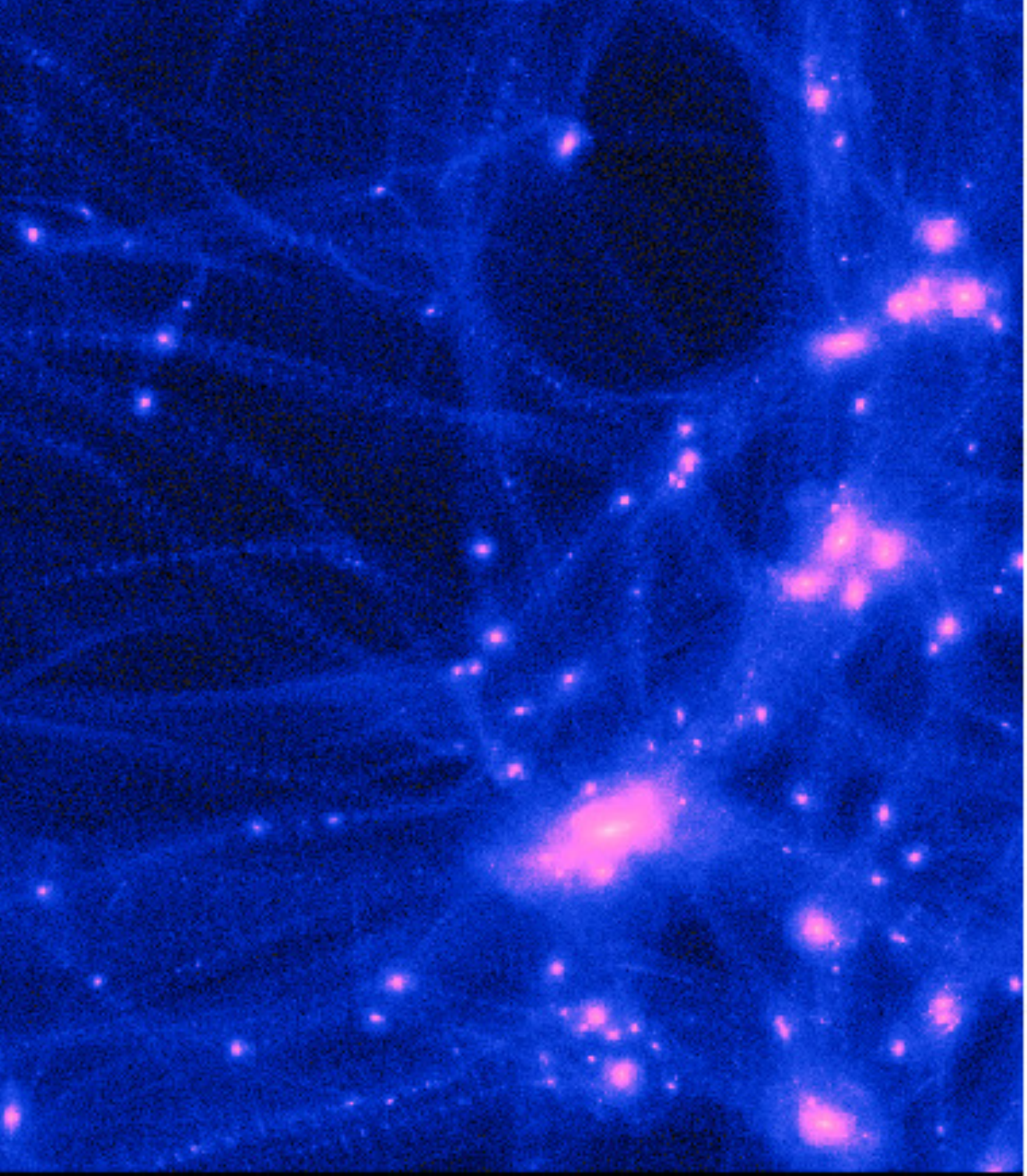,height=290pt} &
\psfig{file=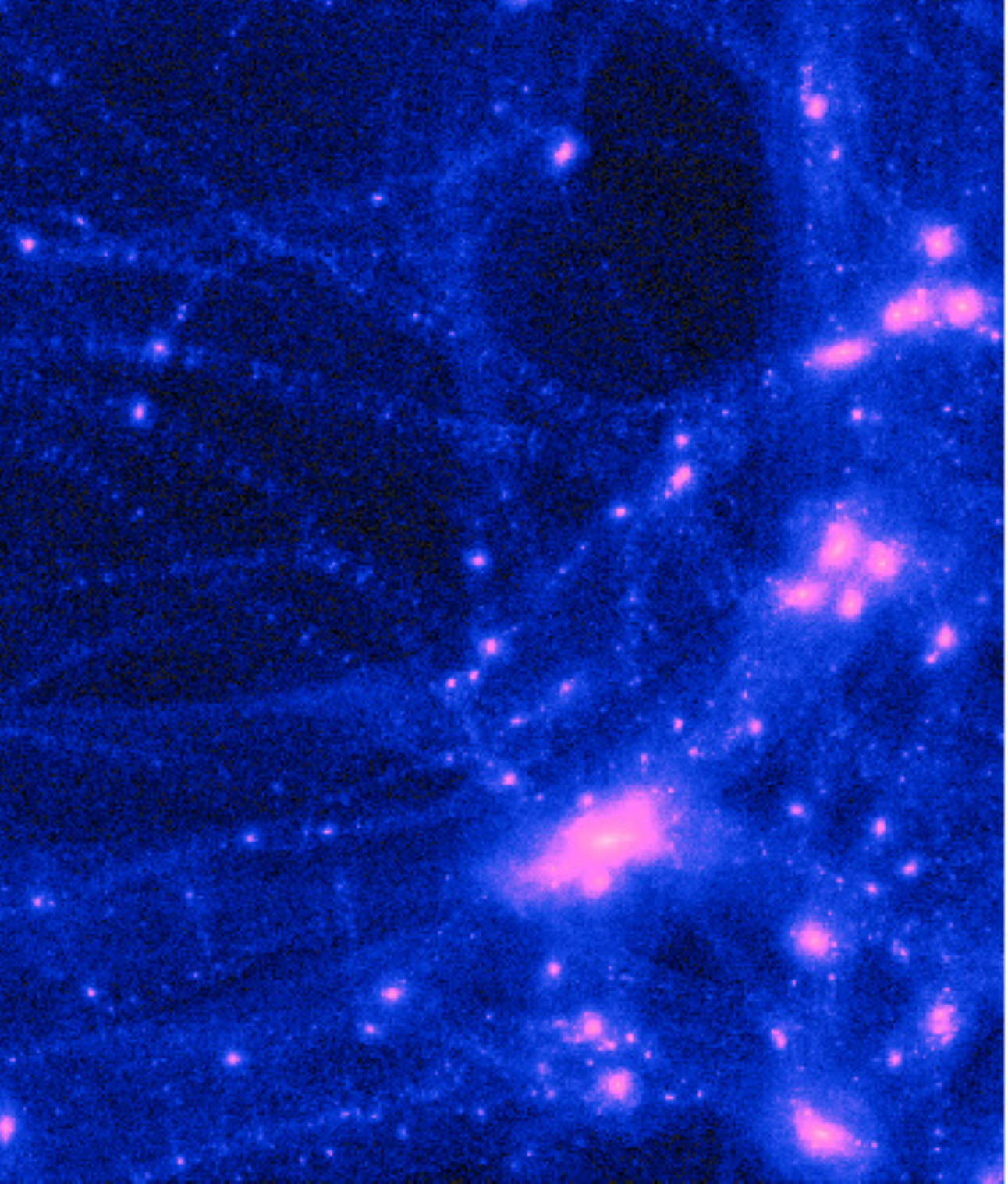,height=290pt} \\
\end{tabular}
\caption{A zoom in simulations with the same cutoff in the power spectrum having corresponding thermal velocities - WDM2(left) and
  no thermal velocities WDM1(right) at z=0 }
\label{fig:200eV} 
\end{figure*}

Due to the free streaming velocity of the particles, the network configuration and architecture of a certain region is rapidly changing. When the density becomes
higher in more isolated regions, the collapse occurs even later, after redshift 4 and some of those halos do not suffer mergers
(Fig.\,\ref{fig:C3}), so there are halos at redshift zero that have formed via a top-down scenario and did not grow through
hierarchical mergers .  This is an interesting result, since the observations show that a large fraction of halos in the universe
have not suffered any mergers until redshift zero.

Why a certain region has more a top-down or bottom-up formation history depends only on the spatial distribution of the
filaments in a certain simulation.

A single halo simulated with different velocities can be seen in the movie halo.avi\footnote{The movie can be found on youtube \url{https://www.youtube.com/watch?v=s_H4dSOP27I}. A zoom in the halo can be watched at \url{https://www.youtube.com/watch?v=zqVi9SSWmXM}}. The $7\times10^{12}\Msun$ halo
forms top-down at the intersection of the filaments and has 18 million particles in its $r_{200}$ radius.

These high resolution runs are 8$^3$ times more resolved in mass than the initial ones: the dark matter particle mass is $m_{d} =
2.72 \times 10^5 \Msun$, where each dark matter particle has a spline gravitation softening of 355\,pc.

Although the WDM halos on galactic scales contain few bound substructures, one can see shells and caustics inside the
virialized region which arise from the coherent infall of material along filaments and from the smooth surrounding
regions. As the resolution increases, the presence of shells and caustics becomes more apparent. The early top-down formation of a halo at the intersection filaments is shown in Fig.\,\ref{fig:halotop} along with a zoom in its central region.  One can clearly see the shells and caustics wrapped inside the 18 million particle halo. A thin slice projection of the warm dark matter halo and a cold dark matter one, clearly illustrates in Fig.\,\ref{fig:slice} how strikingly different is their inner structure.

\subsection{Impact of thermal velocities on structure formation}

As stressed in Section 2, the use of thermal velocities in warm dark matter simulations is crucial, even if their value is
comparable with the Zeldovich velocities at a certain redshift.

In Fig.\,\ref{fig:200eV} we show the differences at redshift zero between the structures emerging in a region in two similar
simulations, WDM1 (without thermal velocities), and WDM2 (with thermal velocities). Both simulations have the same size, the same
power spectrum cutoff and the same initial redshift. The structure formation and evolution in these two simulations is shown side by
side in movie cosmoboxall.avi\footnote{ The movie can be watched here: \url{https://www.youtube.com/watch?v=5txGwBRNC1U}}.

We can see that although the position of the big structures is not affected, below Mpc scales there is a memory of the grid in the
simulation without velocities that is smoothed out when adding thermal velocities, as expected. Some of the very small halos formed
in the simulation without velocities cannot be found in the simulation where thermal velocities are included.  The lack of small
halos in WDM simulations with velocities is of course a crucial feature hinting to resolve the discrepancy between the CDM
simulations predicting too many subhalos in galaxy-sized halos in comparison with the observed number of dwarf galaxies around
large galaxies.  Indeed WDM simulations without velocities still suffer from the infinite phase space density problem.

For comparison, we have performed a suit of simulations that start with a cold distribution of particles, no power spectrum cutoff,
but have velocities corresponding to 1\,km/s, 10\,km/s, 200\,km/s and 700\,km/s. Even the early structure formation is qualitatively
different from the warm dark matter simulations, confirming that the top-down collapse is induced by the damping of the power
spectrum at small scales and not the thermal velocities.

\subsection{Technical aspects in simulating WDM}

The resolution limit poses even a more stringent problem in warm dark matter simulations then in cold dark matter ones. Indeed, in order to properly analyze a region of the simulation, multiple refinements of that region with higher
resolution particles are used. This implies tracking the particle backwards, from redshift zero to the initial
conditions. Due to the large streaming velocities, particles that end up in a virialized halo at redshift zero come from
a larger region than in the CDM simulations, making it more difficult to reach high resolution simulations in WDM.

The heavier the effective mass of our simulation particles, the more prominent is the 2-body relaxation effect in small
clumps (Eq.\,(\ref{eq:relax})).  This problem is more stringent in the case of cold dark matter simulations, where an
initial zero velocity is used. In the case of warm dark matter, this scales with the velocity of the particle, giving a
smaller relaxation time for a smaller velocity. This is why for simulations in the keV range, where the streaming velocity
is smaller, the top-down formation history has been barely observed.

As recently shown by \cite{gao14}, methods like 'FoF' used in analyzing cold dark matter simulations are proved to be insufficient
in analyzing warm dark matter halos.  We confirm this statement, finding that the artificial fragmentation occurring along the
filaments results in a high number of small halos with less than ten particles.

\section{Conclusions and discussion}

We have performed several N-body warm dark matter simulations within a large range of velocity dispersion, for the purpose of
pointing out the effects on the formation of structure. We have then focused on a regime where the resolution is better balanced by
the velocity distribution. Some of our findings are summarized below.
\begin{itemize}
\item In warm dark matter models, as our dark matter only simulations show, the structure formation follows a hybrid scenario in
    which both top-down and bottom-up scenarios have a saying.
\item The early structure formation in this warm dark matter models is essentially top-down, with large halos forming in the
highest density regions, tracked at the intersection of filaments. The second level of top down formation of structure is occurring
along single isolated filaments.
\item The biggest earlier formed halos will accrete matter from the filaments, while in small densities
regions the mergers of smaller halos will result in a larger halo.
\item Later on, and depending on the morphology of the region in which these halos formed, meaning mainly the density
    and the layout of the filaments, they merge into bigger halos creating a hierarchical build-up.
\item The warm dark matter halos, especially the ones that did not suffer big mergers, show obvious shells and caustics.
\item The warmer the dark matter the more pronounced is the top-down effect and the more delayed is the collapse.
\item Albeit the numerical limitations we encounter as far as our warm dark matter simulations are concerned, we can conclude that
  an early top-down structure formation trend would be seen even in dark matter simulations with $v<0.05\, \textrm{ km/s}$. For colder
  particles, this effect is hidden and wiped out by following abundant mergers resulting in a redshift zero distribution that seems
  in agreement with the hierarchical formation scenario.
\item The number of small satellites, as previously found, is visibly reduced in the WDM simulations with respect to the CDM ones.
\end{itemize}

For a warm dark matter particle, as supported by the arguments adduced in Section 2, the thermal component of the velocity is
important for different theoretical and practical considerations.  The strong dependence of the mass-velocity relation on the actual
particle production model makes it difficult to constrain certain properties of the dark matter particle, including its mass. The
impact that a certain velocity dispersion is having on the structure formation and evolution on both small and large scales, as seen
in simulations cannot be used as a strong constraint on the mass in the absence of a universal model for particle production.
Furthermore, we have shown that there have been some inconsistencies in previous studies with respect to the use of velocities in
the simulations, that lead to ambiguous results.
\medskip

The baryonic physics may play an important role in the actual formation and evolution of halos, hence the necessity of further
exploring these effects in simulations. High redshift observations of halos could be used in comparison with complex baryonic warm
dark matter simulations in constraining the mass of warm dark matter particles based on their formation and merger history.

The baryonic processes that we have not included in the simulations must play an important role in the structure
formation. Previously \cite{gao} show a crucial difference in the collapse of a filament that contains both gas and dark matter in a
3\,keV simulation, with respect to the cold dark matter case.  In the WDM case, the stars form inside the filament, before the halo
forms. This trend where stars form in filaments continues for 1.5\,keV particles up to redshift $z\approx2$ resulting in stringy
``chain'' galaxies that remain to be confirmed by observations \citep{gao14}.

The smoother space distribution in the warm dark matter scenario may allow baryons to condense coherently in a smooth potential
halo, providing favorable conditions for forming disk-like galaxies. However, a much higher resolution that the one available in
present simulations is needed to explore this effect.

\section*{Acknowledgments}

S.P. would like especially to thank Doug Potter and Joachim Stadel for their valuable input and help. Discussions with George Lake,
Andrea Macci\`o, and Ben Moore are acknowledged.


\newpage
\appendix
\section{Velocity dispersion dependence on temperature in Fermi-Dirac and Bose-Einstein distributions}
As mentioned in Sect.~\ref{sect:inspect}, we check the correction to velocity dispersion that should be applied to a Maxell-Boltzmann
distribution when the physical system follows a given quantum statistics.
\begin{figure}
\psfig{file=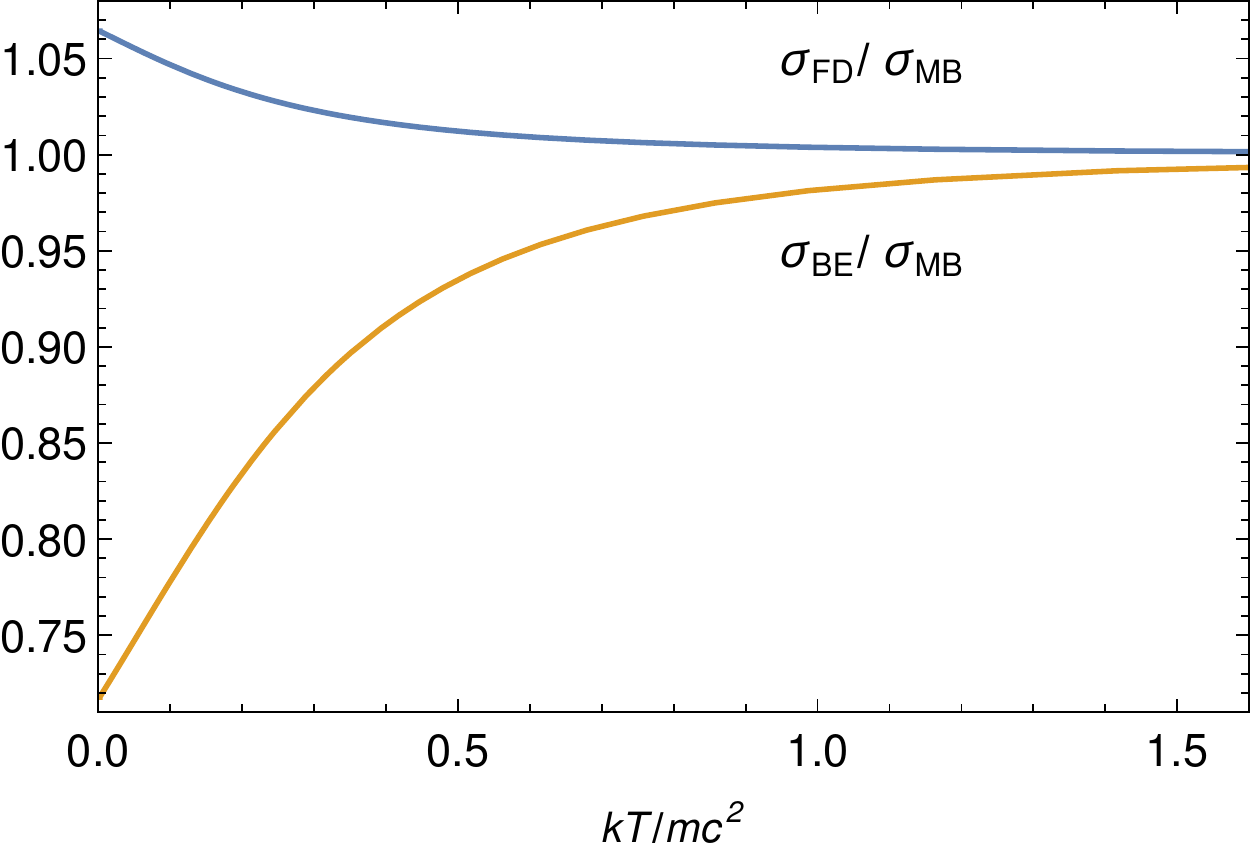, width=230pt} \\
\caption{The ratios between the velocity dispersions $\sigma_{\mathrm{FD}}/\sigma_{\mathrm{MB}}$ and $\sigma_{\mathrm{BE}}/\sigma_{\mathrm{MB}}$ with respect to the temperature}
\label{fig:B1}
\end{figure}

The energy of a particle as a function of momentum $p$, valid in all regimes (relativistic and non-relativistic), is
\begin{equation}
\label{eq:appen}
\epsilon(p)=\sqrt{p^2c^2 + m^2c^4} - mc^2 \ ,
\end{equation}
and the velocity
\begin{equation}
\label{eq:appvel}
v(p)=\frac{p c}{\sqrt{m^2c^2+p^2}} \ .
\end{equation}
The corresponding 1D spherical distributions for Fermi-Dirac, Maxwell-Boltzmann and Bose-Einstein cases are:
\begin{equation}
\label{eq:appFD}
f_{\mathrm{FD}}={4\pi p^2 \over \exp\left(\epsilon(p)/kT\right)+1}
\end{equation}
\begin{equation}
\label{eq:appMB}
f_{\mathrm{MB}}={4\pi p^2 \over \exp\left(\epsilon(p)/kT\right)}
\end{equation}
\begin{equation}
\label{eq:appBE}
f_{\mathrm{BE}}={4\pi p^2\over \exp\left(\epsilon(p)/kT\right)-1}
\end{equation}
Integrating  over all $p$, we obtain the normalization constant
\begin{equation}
S=\int_0^\infty f  dp\ .
\end{equation}
For each case respectively the second moment are 
\begin{equation}
\sigma_{\mathrm{FD}}^2={1 \over S} \int_0^\infty p^2 f_{\mathrm{FD}}\, dp \ ,
\end{equation}
\begin{equation}
\sigma_{\mathrm{MB}}^2={1 \over S} \int_0^\infty p^2 f_{\mathrm{MB}}\, dp \ ,
\end{equation}
\begin{equation}
\sigma_{\mathrm{BE}}^2={1 \over S}\int_0^\infty p^2 f_{\mathrm{BE}}\, dp \ .
\end{equation}
Computing these integrals by numerical quadrature, we find the ratios between the velocity dispersions $\sigma_{\mathrm{FD}}/\sigma_{\mathrm{MB}}$ and
$\sigma_{\mathrm{BE}}/\sigma_{\mathrm{MB}}$ with respect to temperature. The result is plotted in Fig.\,\ref{fig:B1}.  In any situation the
Fermi-Dirac velocity dispersion is not significantly different from Maxwell-Boltzmann's, differing by at most $\sim 6.5\%$, while
the Bose-Einstein velocity dispersion differs more, up to $\sim 27\%$. The highest differences occur at low temperature,
corresponding to low redshifts. This is not such a dramatic correction as the factor 3.571 invoked in \cite{Maccio2012}, but can
still be significant for high precision cosmology works.

\section{Detailed derivation of the results in Section 2.2.2}

Using these expressions inserted into Eq.~(\ref{entropy2}), the specific particle entropy becomes
\begin{equation}
{s\over k} = {1\over3} {\int_0^\infty {y^{3/2}\sqrt{2q+y}  (5q+4y)\over Z^{-1}\exp(y) \pm 1}  \over 
                        \int_0^\infty {y^{1/2}\sqrt{2q+y}  (q+y)  \over Z^{-1}\exp(y) \pm 1}} -\ln(Z) \ ,
\end{equation}
where $Z \equiv \exp(\mu/kT)$, $q\equiv mc^2/kT$ and $y \equiv (\sqrt{p^2c^2+m^2c^4}-mc^2)/kT$.  Thus $s$ is a function of the reduced
dimensionless variables $Z$ and $q$ only, and not of $g$, $m$ and physical constants explicitly.

In the ultra-relativistic regime when the energy of particles is comparable or higher than the rest mass energy, particles and
their antiparticles can be created in equal number, so any chemical potential should cancel to a high degree.  Then $s/k$ at $\mu=0$
becomes a constant.  The closed form expressions are, 
\begin{equation}
\label{entropy2FDUR}
\lim_{T\to\infty} {s(T,0) \over k} = {7\over 135} {\pi^4\over \zeta(3)} \approx 4.20183245,
\end{equation}
for fermions, and 
\begin{equation}
\label{entropy2BEUR}
\lim_{T\to\infty} {s(T,0) \over k} = {4\over 45} {\pi^4\over \zeta(3)} \approx 3.60157071,
\end{equation}
for bosons, where $\zeta$ is Riemann's function, and $\zeta(3)\approx 1.20205690$.

\begin{figure}
\psfig{file=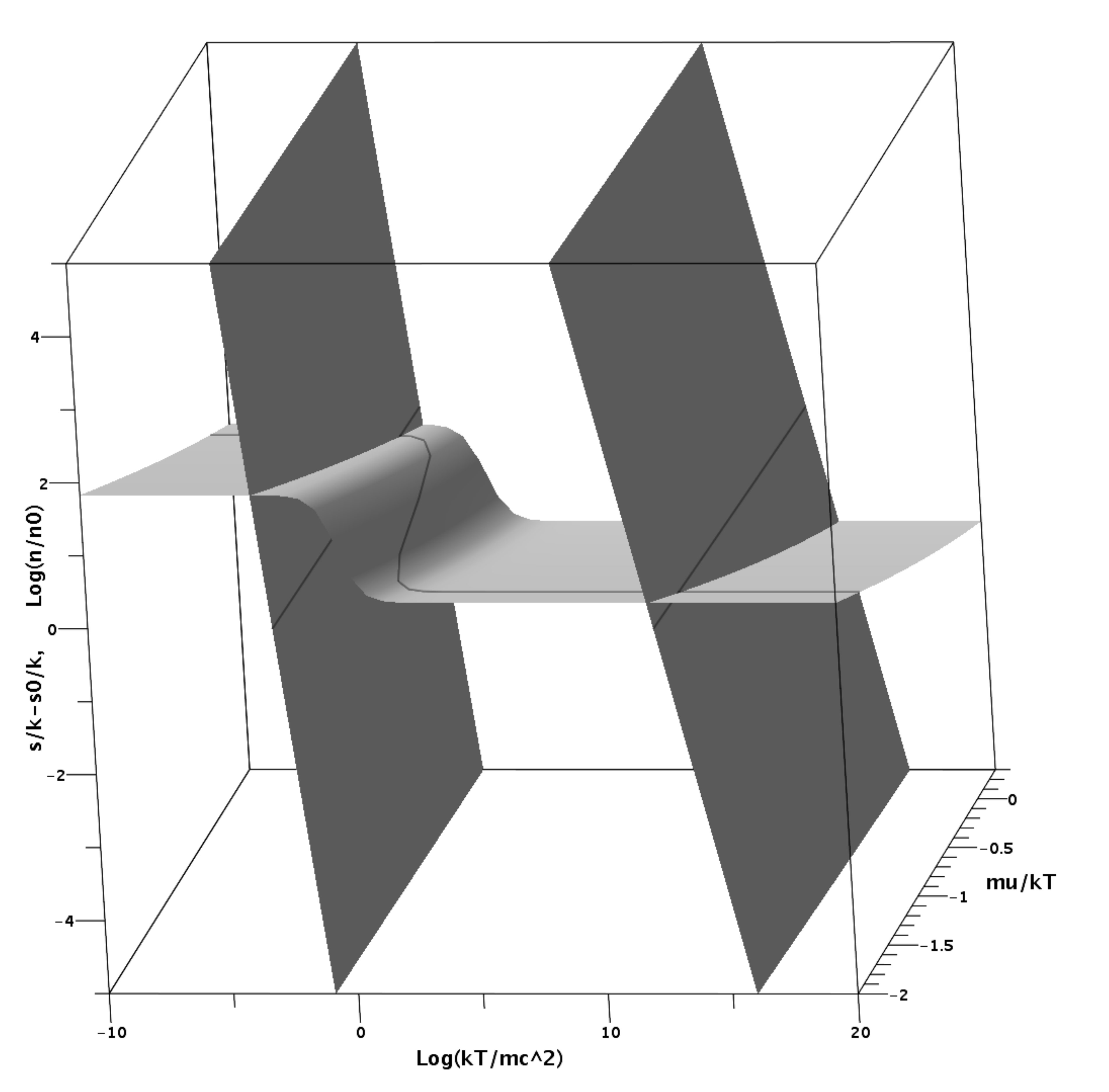, width=250pt} \\
\caption{Density $n$ and entropy $s/k$ as functions of temperature $T$ and chemical potential $\mu$ as given in Eq.~(\ref{nzB}) and
  (\ref{entropyzB}).  The light-grey surface is the entropy and the dark-grey ones are the density at two redshifts $z=10^9$ on the
  left (relativistic), and $z=0$ on the right (non-relativistic), while $n_0=115\,\mathrm{cm}^{-3}$, $g=1$, and
  $m=1 \mathrm{keV}/c^2$. The intersection of the level curves yields the solution of Eq.~(\ref{nzB}) and (\ref{entropyzB}).  }
\label{fig:FCTS}
\end{figure}

The particle velocity at all regimes can be derived from the relativistic particle kinetic energy $\epsilon(T,\mu)=
e(T,\mu)/n(T,\mu)=\sqrt{p^2c^2+m^2c^4}-mc^2$ and that relativistic momentum is related to velocity by $v^2/c^2 = 1 /(1+ m^2 c^2/p^2)$.
Eliminating $p$ yields, noting $Y\equiv \epsilon /mc^2$, 
\begin{eqnarray}
{v^2(T,\mu)\over c^2} &=& 1 - {1\over \left(1+ Y \right)^{2}} = {Y \left( 2 + Y \right)  \over 
                            \left( 1 + Y \right)^2 } \ .
\end{eqnarray}
The second form is numerically more precise at low energy.  
The non-relativistic and ultra-relativistic expansions read, respectively,
\begin{eqnarray} 
  {v^2\over c^2} &\approx& 2Y - 3 Y^2 + 4Y^3- \ldots\\
  {v^2\over c^2} &\approx& 1- Y^{-2} + 2 Y^{-3} - 3Y^{-4} + \ldots
\end{eqnarray}
As stated in Section 2.2.2, the conserved particle density $n(T,\mu)$ is related to universal expansion by the scale factor $a = 1/(1+z)$ and therefore
\begin{equation}
  \label{nzB}
  n(T(z),\mu(z)) = n_0 (1+z)^3 \ ,  
\end{equation}
while the constant particle entropy gives 
\begin{equation}
  \label{entropyzB}
  {s(T(z),\mu(z))\over k} = {s(\infty,0)\over k} = 4.20183245 \ ,  
\end{equation}
For a given particle density $n_0$, redshift $z$ and particle mass $m$ the non-linear Eq.~(\ref{nzB}) and (\ref{entropyzB}) can be
solved with a non-linear equation solver for $T$ and $\mu$.  The functions are univalued and level curves of $n$ and $s$ intersect
once, so any combination of $T$ and $\mu$ gives a single solution (see Fig.~\ref{fig:FCTS}).  Actually the constant level curves
$n$ and $s$ expressed with the variables $\log q$ and $Z$ intersect almost at right angle: $n(\log q,Z)$ depends most rapidly on
$\log q$, and $s(\log q,Z)$ depends most rapidly on $Z$, so finding a solution for $\log q$ for $n$ at constant $Z$ and then a
solution for $Z$ at constant $q$ for $s$, and repeating until satisfaction could be a method to find a solution.  Since the
thermodynamic functions involve integrals, a fast numerical integrator is handy, since several indefinite integrals must be evaluated
at each iteration.  To perform this we used Maple 18 which includes a non-linear multidimensional function root solver, and evaluate
quickly numerical integrals with the NAG library algorithm D01AMC.\footnote{The Maple script is available on request.}  When $T$ and
$\mu$ are found for a given particle mass and degeneracy factor $g$, all the other quantities like $v^2$ can be derived by plugging
these values in the functions, which may require again few numerical integral evaluations.  The results are presented in Section
2.2.2.

\section{Movies captions and snapshots}

\begin{table*}
\floatplacement{figure}{H}
\caption{Description of the movies accompanying the paper}
\label{table:movies}
\begin{tabular}{@{}lc}
\hline
Label & Description\\ 
\hline
cosmoboxvel.avi & Movie of full-box WDM2 simulation\\
cosmoboxall.avi & WDM1 and WDM2 full-box simulations side-by-side showing the effect of thermal velocities\\
lu.avi ld.avi ru.avi rd.avi & A zoom in the $1/4$ of the WDM2 simulation \\
halo.avi & A $7\times 10 ^{12} M_{\sun} $ $18\times 10^{6}$  particles high-resolution refined halo from WDM5\\
halozoom.avi & A zoom in the refined halo focused on the central region where the shells and caustics can be observed\\
\hline
\end{tabular}
\end{table*}
\begin{figure*}
\psfig{file=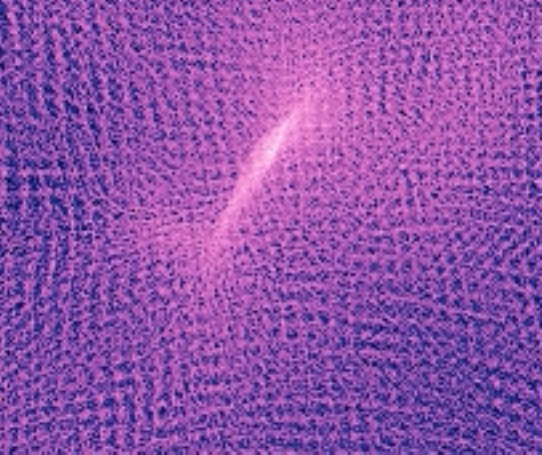,height=105pt, width=105pt} \ 
\psfig{file=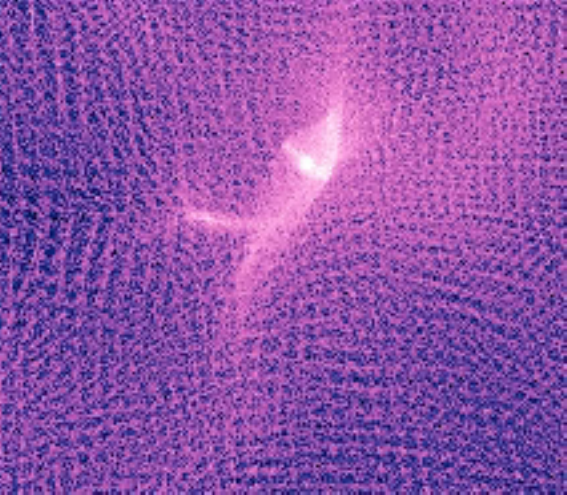,height=105pt, width=105pt} \
\psfig{file=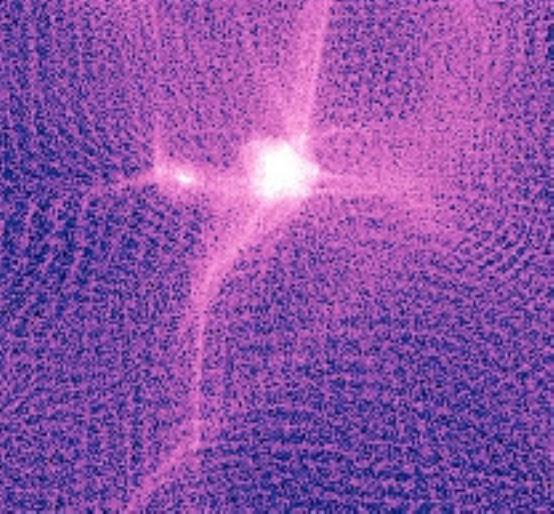,height=105pt, width=105pt} \
\psfig{file=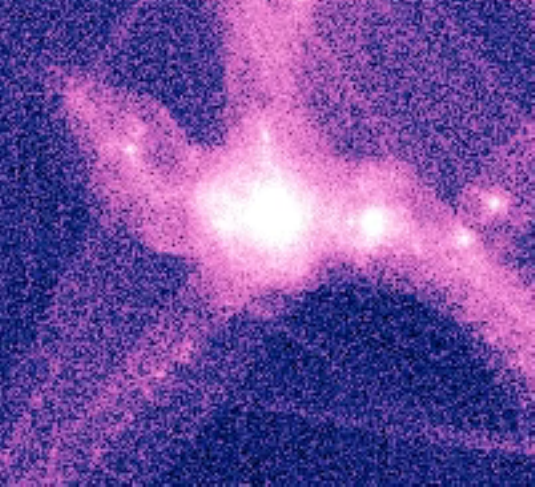,height=105pt, width=105pt} 
\caption{A zoom in a region from the WDM2 simulation, showing the evolution of a halo which forms top-down at the intersection of the filaments and then starts accreting matter}
\label{fig:C1}
\end{figure*}
\begin{figure*}
\psfig{file=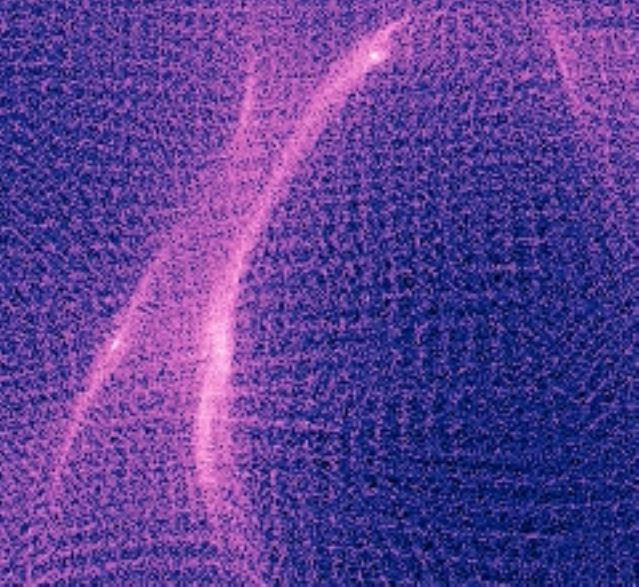,height=105pt, width=105pt} \
\psfig{file=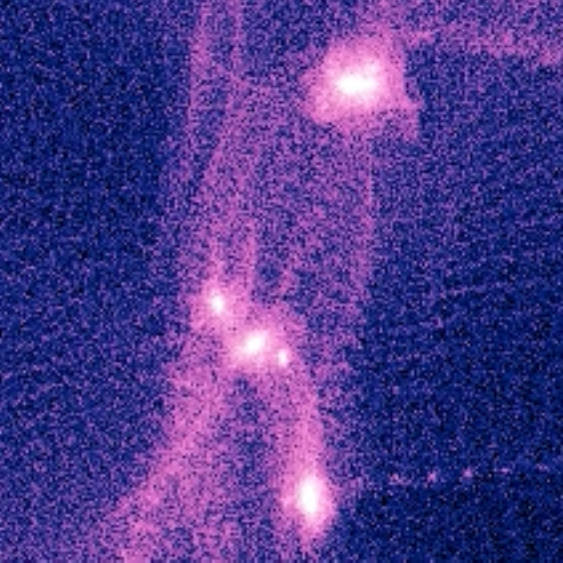,height=105pt, width=105pt} \ 
\psfig{file=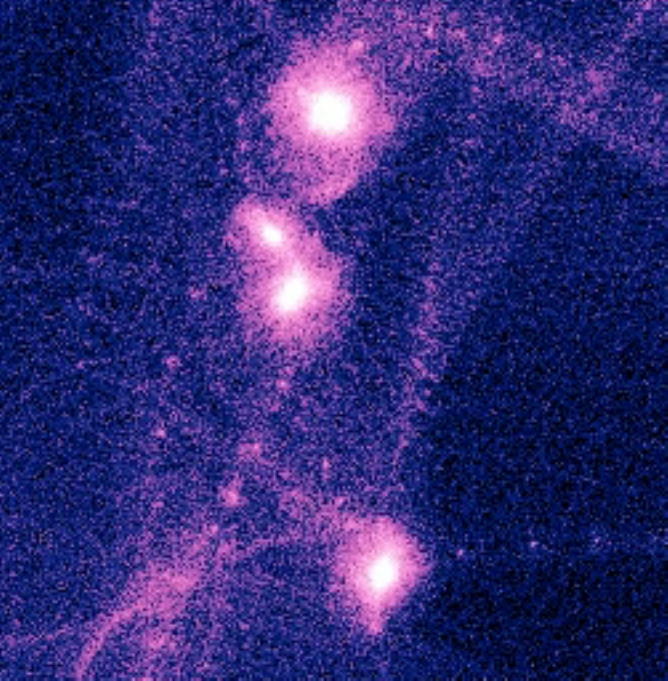,height=105pt, width=105pt} \ 
\psfig{file=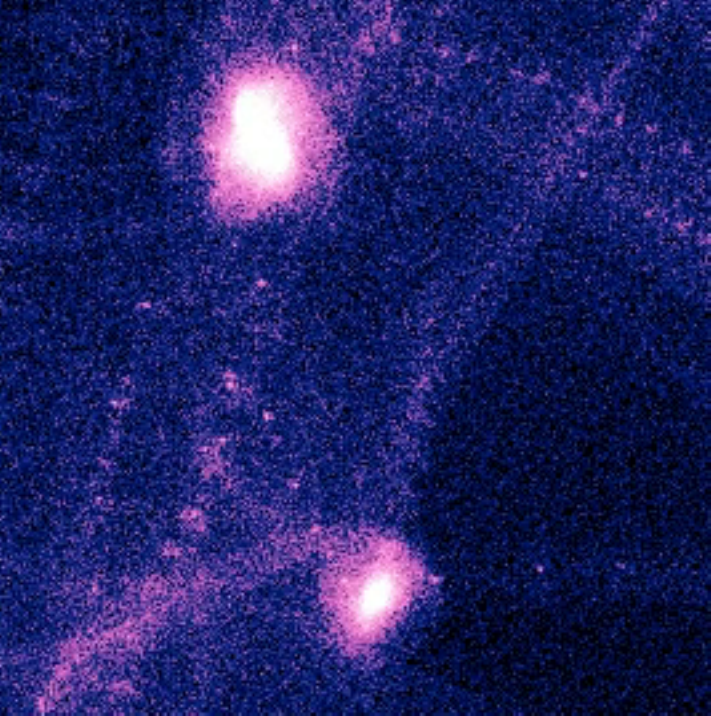,height=105pt, width=105pt} 
\caption{A zoom in a region from the WDM2 simulation showing how small halos formed later that merge hierarchically in a larger halo}
\label{fig:C2}
\end{figure*}
\begin{figure*}
\psfig{file=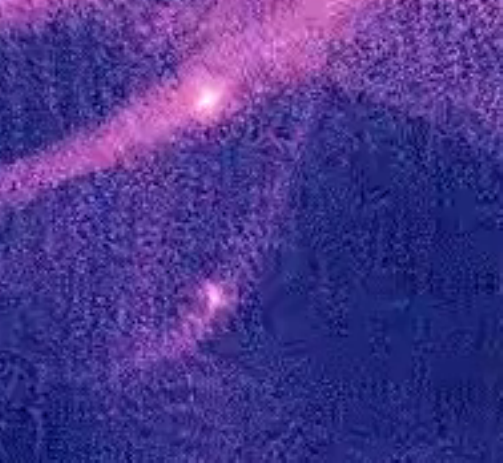,height=105pt, width=105pt} \ 
\psfig{file=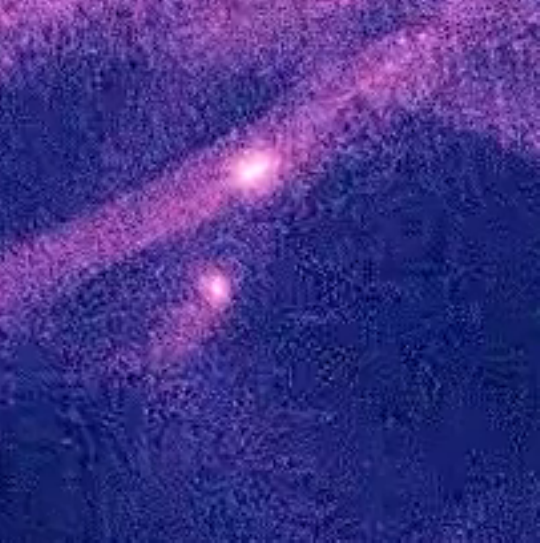,height=105pt, width=105pt} \ 
\psfig{file=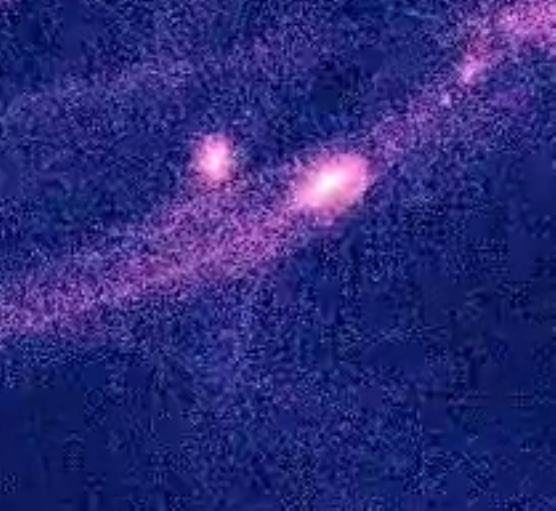,height=105pt, width=105pt} \ 
\psfig{file=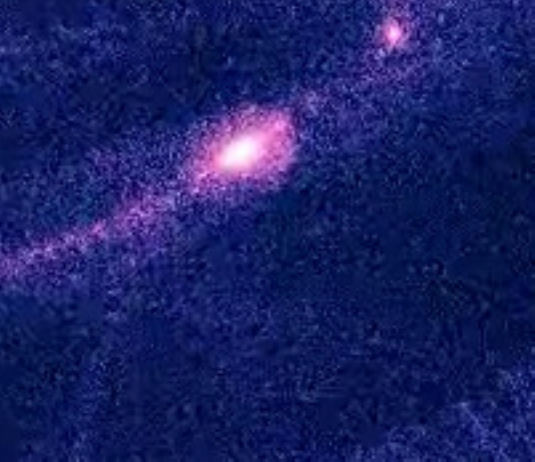,height=105pt, width=105pt} 
\caption{An early formed halo which doesn't suffer mergers}
\label{fig:C3}
\end{figure*}
\begin{figure*}
\psfig{file=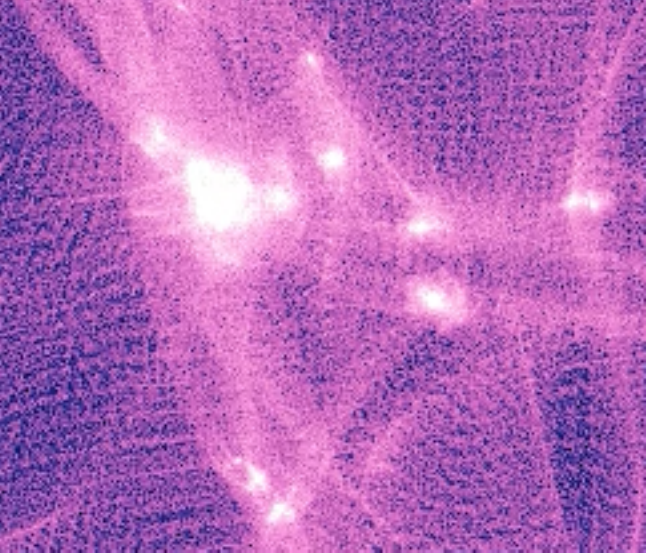,height=105pt, width=105pt} \ 
\psfig{file=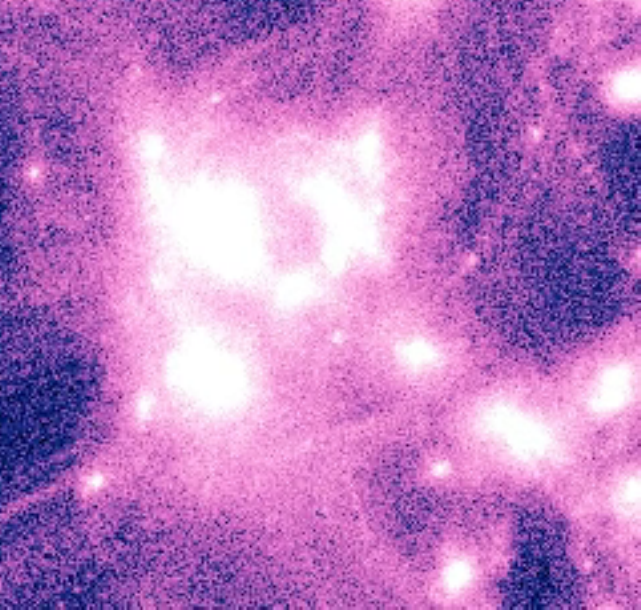,height=105pt, width=105pt} \ 
\psfig{file=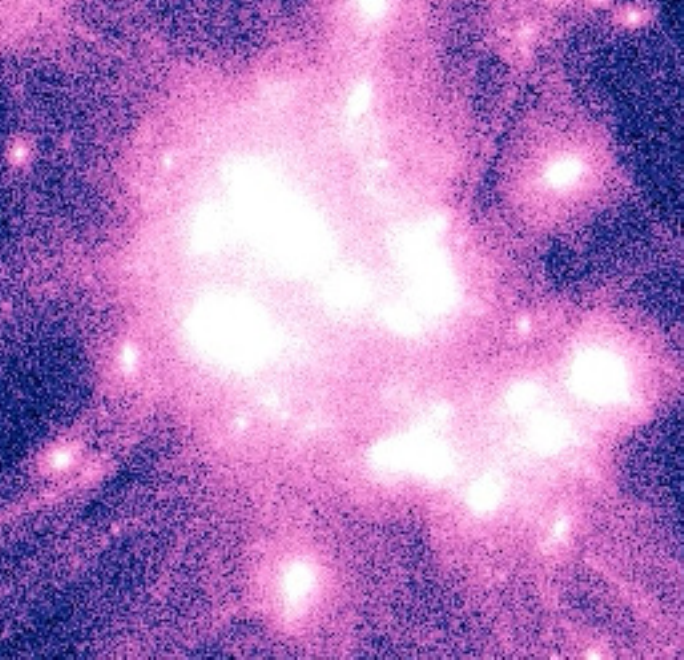,height=105pt, width=105pt} \ 
\psfig{file=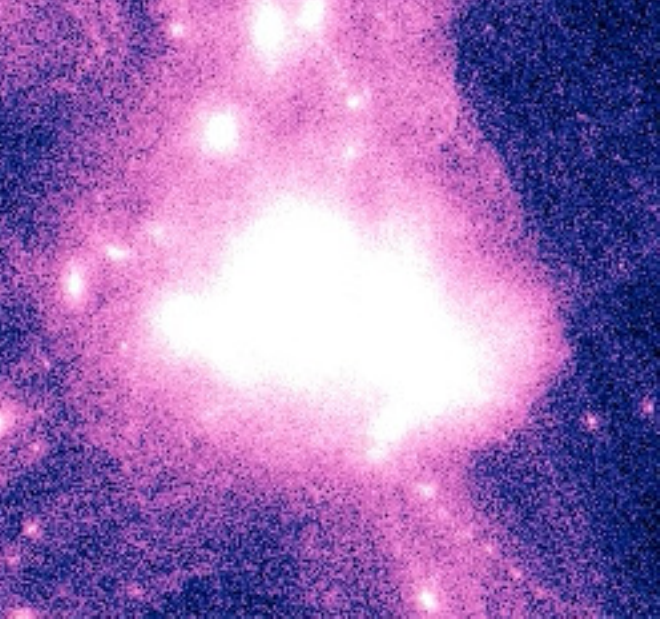,height=105pt, width=105pt} 
\caption{A large high density region with many filaments where the halos formed early on via top-down collapse are merging in a violent manner creating a larger cluster}
\label{fig:C4}
\end{figure*}
\end{document}